\documentclass[english]{article}
\usepackage{jheppub}
\usepackage{amsthm}
\usepackage{epstopdf}
\usepackage[ruled,vlined,linesnumbered]{algorithm2e}
\providecommand{\definitionname}{Definition}
\newtheorem{defi}{\protect\definitionname}
\providecommand{\claimname}{Claim}
\newtheorem{claim}{\protect\claimname}
\providecommand{\propname}{Proposition}
\newtheorem{prop}{\protect\propname}
\newcommand{\matr}{\mathbb}
\newcommand{\diag}{\mathop\mathrm{diag}}
\newcommand{\bal}{\mathcal{B}}
\allowdisplaybreaks

\title{Reducing differential equations for multiloop master integrals}

\author{Roman N. Lee}
\affiliation{ Budker Institute of Nuclear Physics, 630090, Novosibirsk, Russia}
\emailAdd{r.n.lee@inp.nsk.su}
\abstract{
We present an algorithm of the reduction of the differential equations for master integrals the Fuchsian form with the right-hand side matrix linearly depending on dimensional regularization parameter $\epsilon$. We consider linear transformations of the functions column which are rational in the variable and in $\epsilon$. Apart from some degenerate cases described below, the algorithm allows one to obtain the required transformation or to ascertain irreducibility to the form required. Degenerate cases are quite anticipated and likely to correspond to irreducible systems.
}
\begin{document}
\maketitle
\flushbottom
\section{Introduction}

For a few last decades, the demand for the multiloop calculations is constantly growing, the methods of such calculations evolved accordingly. For multiscale integrals, probably, the most powerful technique is the differential equations method \cite{Kotikov1991,Kotikov1991a,Kotikov1991b,Remiddi1997,GehrRem2000}. 
Within this method, the master integrals are found as solutions of the differential equations obtained with the help of the IBP reduction \cite{Tkachov1981,ChetTka1981,Laporta2000}.

Recently, a remarkable observation has been made by Henn in Ref. \cite{Henn2013} concerning the differential equations method. Namely, it appeared that in many cases the dependence on the dimensional regularization parameter $\epsilon$ of the right-hand side of the differential equations for the master integrals can be reduced to a single factor $\epsilon$ by a judicious choice of the master integrals. For brevity in what follows we will refer to such a form of the differential system as \emph{$\epsilon$-form}. With this form (and also the initial conditions) at hand, finding the solution up to any fixed order in $\epsilon$ becomes a trivial task. Moreover, the solution manifestly possesses a remarkable property of homogeneous transcendental weight. Since then a number of papers successfully applied this approach to the calculation of various classes of integrals \cite{Henn:2013tua,HennSmirnov2013,HennSmirnovSmirnov2014,Caron-HuotHenn2014,GehrmannManteuffelTancrediWeihs2014,GrozinHennKorchemskyMarquard2014,DiVita:2014pza,Hoschele:2014qsa,LiManteuffelSchabingerZhu2014,ManteuffelSchabingerZhu2014,BellHuber2014}.

In general, finding an appropriate basis is not easy. In Ref. \cite{Henn2013} two guiding principles have been suggested. The first method is based on the examination of generalized unitarity cuts, and the second one is based on finding integral $d\log$ form. Both  methods may be used (with some amount of heuristic work) for determining whether a specific integral is homogeneous or not, however, in general, they do not give an algorithm of finding appropriate basis (though, they proved their validity in a number of applications). In Refs. \cite{ArgeriDiMastroliaMirabellaSchlenkothers2014,GehrmannManteuffelTancrediWeihs2014} algorithms of the reduction have been presented assuming a very special form of the differential system. Despite these advances, finding an appropriate basis has been rather an art than a skill so far.  Therefore, devising a practical algorithm of finding the described form of the differential system is of essential interest.
 
In the present paper we describe a method of finding an appropriate basis which is based on the differential system alone. 
The system can be written in the matrix form 
\begin{equation}\label{eq:DEsystem}
\partial_{x}\mathbf{J}=\matr{M}\left(\epsilon,x\right)\mathbf{J}\,,
\end{equation}
where $\epsilon$ is the dimensional regularization parameter ($d=4-2\epsilon$), $x$ is some parameter, $\mathbf{J}$ is the column of the master integrals, $\matr{M}$ is $n\times n$ matrix, rational in both $\epsilon$ and $x$. 

Our main algorithm can be divided into three stages. At first stage the differential system is reduced to the Fuchsian form, i.e., to a form when the elements of $\matr{M}$ have only simple poles with respect to $x$. After this stage, the matrix can be written as 
\begin{equation}
\matr{M}\left(\epsilon,x\right)=\sum_k \frac{\matr{M}_k(\epsilon)}{x-x_k}\,.
\end{equation}
Note that this step is always doable for the systems with regular singularities. Possibility to reduce the system to Fuchsian form is known since works \cite{Roehrl1962,Roehrl1957} of R\"ohrl and the specific algorithm for this reduction can be easily deduced from that of Barkatou\&Pfl\"ugel \cite{BarkatouPfluegel2009,BarkatouPfluegel2007}, see below. Algorithm \ref{alg:Fuchsian} of the present paper is advantageous only in that it tries to minimize the number of apparent singularities generated during the reduction process. At second stage the eigenvalues of $\matr{M}_k$ are normalized, i.e., their real parts are reduced to the interval $[-1/2,1/2)$. For  the systems reducible to $\epsilon$-form this means that all eigenvalues are made proportional to $\epsilon$. It is easy to see that, when this step is successful, the resulting system has no apparent singularities, see Eq. \eqref{eq:monodromy} and discussion after it. Finally, a constant transformation is searched for in order to factor out $\epsilon$, i.e., to reduce the system to $\epsilon$-form. We give one nontrivial example of the application of our algorithm.

Except for the last stage, our algorithm is not specific to the systems depending on parameter. In particular, it can be used to eliminate apparent singularities and to find the matrices of monodromy around singular points (up to similarity).

\section{Preliminaries}

We consider the system of differential equations for the master integrals as given in Eq. \eqref{eq:DEsystem}.
Under the change of functions 
\begin{equation}\label{eq:FunctionsChange}
\mathbf{J}= \matr{T}\left(\epsilon,x\right) \widetilde{\mathbf{J}}
\end{equation}
the system modifies to an \emph{equivalent} system
\begin{equation}\label{eq:DEsystem1}
\partial_{x}\widetilde{\mathbf{J}}=\widetilde{\matr{M}}\left(\epsilon,x\right)\widetilde{\mathbf{J}}\,,
\end{equation}
where 
\begin{equation}\label{eq:MatrixTransformationLaw}
\widetilde{\matr{M}}=\matr{T}^{-1}\matr{M}\matr{T}-\matr{T}^{-1}\partial_{x}\matr{T}\,.
\end{equation}

The observation of Ref. \cite{Henn2013} states that it is often possible to find a transformation $\matr{T}$ so that the new column $\widetilde{\mathbf{J}}$ satisfies a simple equation
\begin{equation}
\partial_{x}\widetilde{\mathbf{J}}=\epsilon\matr{S}(x)\widetilde{\mathbf{J}}\,.
\label{eq:CanonicalForm}
\end{equation}
Though it is not stated explicitly in Ref. \cite{Henn2013}, we will require that the matrix $\matr{S}$ has a Fuchsian form, i.e.,
\begin{equation}
\matr{S}(x)=\sum_{k}\frac{\matr{S}_k}{x-x_k}\,,
\label{eq:FuchsianMatrix}
\end{equation}
where $k$ runs over finite set.
This condition is very important on its own because the form \eqref{eq:FuchsianMatrix} allows one to express the result in terms of generalized harmonic polylogarithms. In what follows we will often omit $\epsilon$ in the arguments of functions unless it may lead to confusion.

\begin{defi}
The differential system \eqref{eq:DEsystem}
is said to have a  \textbf{regular singularity} at $x=x_0\neq\infty$ (at $x=x_0=\infty$) if $x=x_0\neq\infty$ is a singular point of $\matr M(x)$ ($y=0$ is a singular point of $M(1/y)/y^2$) and all solutions of the system grow at most like a finite power of $x-x_0$ (of $x$) in the sectorial vicinity of $x_0$. 
\end{defi}

The power-like growth of the master integrals (which are the solutions of the system) in the vicinity of any point follows from their parametric representation. Therefore, it is natural to expect that all singular points of the differential system for the master integrals are regular singularities. 

An \emph{apparent singularity} is a regular singularity which is a finite-order pole or a regular point of any solution of the system. Therefore, the monodromy around an apparent singularity is an identity matrix. As we shall see, it means that, locally, we can always remove apparent singularity with a rational transformation.

\begin{defi}
The differential system \eqref{eq:DEsystem} is said to have  \textbf{Poincar\'e rank} $p\geqslant0$ at the singular point $x=x_0\neq\infty$ if $\matr{M}(x)$ can be represented as $\matr{M}(x)= \matr{A}(x-x_0)/(x-x_0)^{1+p}$, where $\matr{A}(x)$ is regular at $x=x_0$ matrix and $\matr{A}(0)\neq 0$. The system is said to have Poincar\'e rank $p\geqslant0$ at the point $x=\infty$ if $\matr{M}(x)$ can be represented as $\matr{M}(x)=\matr{A}(1/x)x^{-1+p}$, where $\matr{A}(y)$ is a regular at $y=0$ matrix and $\matr{A}(0)\neq 0$. 

If $p=0$, we say that the system is \textbf{Fuchsian} in $x=x_0$ and call $\matr{A}(0)$ a \textbf{matrix residue}. Respectively, we call $x_0$ a Fuchsian point of the system.
\end{defi}

It is easy to show that when the Poincar\'e rank of a system is zero at some point, this point is a regular singularity of the system. But the converse is not always true. However, if some point is a regular singularity, it is possible \emph{to transform} the system to the equivalent one with zero Poincar\'e rank at that point. More generally, Moser \cite{Moser1960} has given necessary and sufficient condition of the possibility to reduce the (generalized) Poincar\'e rank of the system and also presented an algorithm for finding the appropriate transformation matrix. Barkatou and Pfl\"ugel have given an improved version of the algorithm in Refs. \cite{BarkatouPfluegel2007,BarkatouPfluegel2009}. Their algorithm consists of a sequence of rational transformations, each lowering the generalized Poincar\'e rank $p+r/n-1$, where $r=\mathop{\rm rank}\matr{A}(0)$ and $n$ is the size of $\matr{A}(0)$. Applying these transformations several times for each singularity, one can minimize the Poincar\'e rank of all singularities, except maybe one (usually chosen to be $x=\infty$). In particular, if all singularities are regular, after the application of the algorithm, Poincar\'e ranks for all but one singularities can be nullified and thus the system is reduced to a Fuchsian form everywhere, except, may be, one point. In fact, their algorithm also allows one to transform a regular system to Fuchsian form globally with a penalty of introducing some apparent singularities.

The possibility to transform a regular system to Fuchsian form in all points and to eliminate all apparent singularities would mean the positive solution of the 21st Hilbert problem, consisting of proving of the existence of linear differential equations having a prescribed monodromy group. However, Bolibrukh in Ref. \cite{Bolibrukh1989} has proved by presenting an explicit counterexample, that it is not always possible and thus 21st Hilbert problem has negative solution. Nevertheless, the problem of reducing, when it is possible, a rational differential system to Fuchsian form without apparent singularities is very important. An ultimate solution of this problem in the most general case, and, in particular, deciding whether such a reduction is possible, is not known so far to the best of our knowledge.

\begin{defi}
The transformation \eqref{eq:MatrixTransformationLaw} generated by the matrix $\matr{T}(x)$ \textbf{is regular} at $x=x_0\neq\infty$ (at $x=\infty$) if $\matr{T}(x)=\matr{T}_0+O(x-x_0)$ ($\matr{T}(x)=\matr{T}_0+O(1/x)$) and $\det\matr{T}_0\neq0$.
\end{defi}
In this definition the condition $\det\matr{T}_0\neq0$ simply states that $\matr{T}^{-1}(x)$ is also a power series near the point $x=x_0$ ($x=\infty$). Naturally, regular transformations can not change the pole order of $\matr M$, so we have to consider singular transformations. While there are transformations singular at only one point on the extended complex plane, their form appears to be too restrictive for our purposes\footnote{See, however Section \ref{sec:btform}.}. The key tool of our approach is the transformation singular at two points. \begin{defi}
A \textbf{balance} is a transformation, generated by the matrix $\matr{T}$ of the form
\begin{equation}\label{eq:balance}
\matr{T}(x)=\matr{\bal}(\matr{P},x_1,x_2|x)\stackrel{\mathrm{def}}{=}\overline{\matr{P}}+c\frac{x-x_2}{x-x_1}\matr{P}\,,
\end{equation}
where $c$ is some constant, $\matr{P},\ \overline{\matr{P}}$ are the two complementary projectors, i.e. $\matr{P}^2=\matr{P}$ and $\overline{\matr{P}}=\matr{I}-\matr{P}$. More specific, we call the transformation generated by \eqref{eq:balance} the $\matr{P}$-balance between $x_1$ and $x_2$.
\end{defi}
Note that this transformation appears in the consideration of the Riemann problem in complex analysis, see, e.g., Ref. \cite{zakharov1980soliton}. We will always put $c=1$ when both $x_1$ and $x_2$ are finite. When $x_1=\infty$ (when $x_2=\infty$), we put $c=x_1$ ($c=1/x_2$) and understand $c(x-x_2)/(x-x_1)$ as a limit for $x_1\to\infty$ (for $x_2\to\infty$). 

The inverse of the balance is also a balance, since 
\begin{equation}
\matr{\bal}(\matr{P},x_1,x_2|x)\matr{\bal}(\matr{P},x_2,x_1|x)=\matr{I}\,,
\end{equation}
Therefore, the transformation \eqref{eq:balance} is regular everywhere, except the points $x=x_1$ and $x=x_2$, where, respectively, $\matr{T}(x)$ and $\matr{T}^{-1}(x)$ have simple poles.

\section{Reduction at one point}
The basic idea of reducing the Poincar\'e rank is to find such a projector $\matr{P}$ that the transformation generated by \eqref{eq:balance} lowers the rank of $\matr{A}_0$. For a regular singularity, the idea is to use \eqref{eq:balance} to normalize the eigenvalues of the matrix residue.

Let us concentrate on the reduction of the differential system at one point. Without loss of generality, we assume that $x=0$ is a singular point of the system \eqref{eq:DEsystem} and the Laurent series expansion of $\matr{M}(x)$ near $x=0$ has the form
\begin{equation}\label{eq:Mseries0}
\matr{M}(x)=\matr{A}_0x^{-p-1}+\matr{A}_1x^{-p}+O(x^{-p+1})\,.
\end{equation}

\subsection*{Lowering Poincar\'e rank}

First, let us consider the problem of lowering of the Poincar\'e rank, so $p>0$ in this subsection. We assume that $\matr{A}_0$ is a nilpotent matrix since it is a necessary condition for the existence of a transformation which lowers the Poincar\'e rank \cite{Moser1960}. Therefore, $\matr{A}_0$ can be reduced to Jordan form with zero diagonal. Let $r=\mathop{\rm rank}\matr{A}_0$, then a necessary and sufficient condition of existence of a transformation lowering the generalized Poincar\'e rank $p+r/n-1$ introduced in Ref.  \cite{Moser1960} is that
\begin{equation}
\left.x^r \det(\matr{A}_0/x+\matr{A}_1-\lambda \matr{I})\right|_{x=0}=0\label{eq:crit1}
\end{equation}
identically as a function of $\lambda$. 

It is convenient to use an equivalent form of this condition, which was  introduced in Ref. \cite{BarkatouPfluegel2007}. Let $\{u_k^{(\alpha)}|k=1\ldots N,\alpha=0,\ldots n_k\}$ be a basis constructed of the generalized eigenvectors of $\matr{A}_0$ with the properties
\begin{equation}
\matr{A}_0 u_k^{(0)}=0\,,\quad \matr{A}_0 u_k^{(\alpha+1)}=u_k^{(\alpha)}.
\label{eq:uproperties}
\end{equation}
Here $N$ is a number of Jordan cells (including the trivial ones), $n_k$ is a rank of $k$-th Jordan cell, which is its dimension minus one. In what follows we assume that Jordan cells are ordered by their sizes, so that $n_1\geqslant n_2\geqslant\ldots\geqslant n_N$.
Let 
\begin{equation}\label{eq:u_viaU}
\matr U=\left(u_1^{(0)},\ldots ,u_1^{(n_1)},u_2^{(0)},\ldots,u_2^{(n_2)},\ldots\right)
\end{equation} 
be the matrix with columns $u_k^{(\alpha)}$.  This matrix generates the similarity transformation $\matr{A}_0\to \widetilde{\matr{A}}_0=\matr U^{-1}\matr{A}_0\matr U$ reducing  $\matr{A}_0$ to Jordan form. Then 
\begin{equation}\label{eq:v_viaU}
\matr U^{-1}=(v_1^{(n_1)},\ldots ,v_1^{(0)},v_2^{(n_2)},\ldots,v_2^{(0)},\ldots)^{\dagger}\,,
\end{equation}
where $v_k^{(\alpha)}$ are the generalized eigenvectors of $\matr A^{\dagger}_0$ satisfying
\begin{equation}
v_k^{(0)\dagger}\matr{A}_0 =0\,,\quad v_k^{(\alpha+1)\dagger}\matr{A}_0 =v_k^{(\alpha)\dagger}\,.
\label{eq:vproperties}
\end{equation}
We will call $v_k^{(\alpha)\dagger}$ the \emph{left} generalized eigenvectors of $\matr A_0$, in contrast to $u_k^{(\alpha)}$ which we will call the \emph{right} generalized eigenvectors of $\matr A_0$.

From $\matr U^{-1}\matr U=\matr I$ we have
\begin{equation}
v_k^{(\alpha)\dagger}u_l^{(\beta)}=\delta_{kl} \delta_{\alpha+\beta,n_k}\,, 
\label{eq:uvorthonormality}
\end{equation}
so that $\{u_k^{(\alpha)}|k=1,\ldots, N;\alpha=0,\ldots, n_k\}$ and $\{v_k^{(\alpha)}|k=1,\ldots, N;\alpha=n_k,\ldots, 0\}$ are the dual bases.

One observes that relations \eqref{eq:uproperties}, \eqref{eq:vproperties}, \eqref{eq:uvorthonormality} are invariant under the following basis transformation:
\begin{equation}\label{eq:JFSymmetries}
u_k^{(\alpha)}\to u_k^{(\alpha)}+c u_l^{(\alpha)}\,,\quad v_l^{(n_l-\alpha)}\to v_l^{(n_l-\alpha)}-c v_k^{(n_k-\alpha)},\quad (\alpha=0,1,\ldots n_k)\,,
\end{equation}
where $c$ is an arbitrary number, and $k$ and $l$ are some fixed Jordan cell numbers, $k>l$ (we remind that $n_1\geqslant n_2\geqslant\ldots\geqslant n_N$ in our convention).

The above transformation corresponds to the transformation of the matrix $\matr{U}$:
\begin{equation}\label{eq:U_transformations}
\quad\matr{U}\to \matr{U}(\matr{I}+c \matr{E}^{(l,k)})\,,
\end{equation} 
where $(\matr{E}^{(l,k)})_{\widehat{i\alpha}\widehat{j\beta}}=\delta_{il}\delta_{jk}\delta_{\alpha\beta}$.
Here we denoted by $\widehat{k\alpha}$ the number of the column in which $u_k^{(\alpha)}$ stands in $\matr{U}$.  
The condition  \eqref{eq:crit1} can be written as \cite{BarkatouPfluegel2007,BarkatouPfluegel2009}
\begin{equation}\label{eq:BF_condition}
\det \matr{L}(\lambda)=\det (\matr{L}_0+\lambda \matr{L}_1)=0\,,
\end{equation}
where
\begin{equation}
\matr{L}(\lambda)=\matr{L}_0+\lambda \matr{L}_1=[v_k^{(0)\dagger}(\matr{A}_1+\lambda \matr{I})u_l^{(0)}]\,\quad (k,l=1\ldots N).
\end{equation}

The transformation \eqref{eq:JFSymmetries} induces the following transformation of the matrix $\matr{L}_0$:
\begin{equation}\label{eq:L_transformations}
\quad\matr{L}_0\to (\matr{I}-c\delta_{n_kn_l}{\Delta}^{(l,k)})\matr{L}_0(\matr{I}+c \Delta^{(l,k)})\,,
\end{equation}
where $\Delta^{(l,k)}$ is the matrix with unity on the intersection of $l$-th row and $k$-th column and zero elsewhere, i.e. ${\Delta}^{(l,k)}_{ij}=\delta_{il}\delta_{jk}$. It is easy to check that $\matr{L}_1$ is invariant under these transformations. General composition of the transformations of the form \eqref{eq:U_transformations} can be written as
\begin{gather}
\quad\matr{U}\to \matr{U}(\matr{I}+\matr E)\,,\label{eq:U_general_transformation}\\
\quad\matr{L}_0\to (\matr{I}-\widetilde\Delta)\matr{L}_0(\matr{I}+\Delta)\,,\label{eq:L_general_transformation}\\
\matr E=\sum_{l,k;\,l<k}c_{l,k} \matr{E}^{(l,k)}\,,\quad \Delta=\sum_{l,k;\,l<k}c_{l,k}{\Delta}^{(l,k)}\label{eq:EDelta_general}
\end{gather}
The expression for $\widetilde{\Delta}$ can be derived from the representation $\matr{I}+\Delta=\prod (\matr{I}+c_i\Delta^{(l_i,k_i)})$, but its explicit form is irrelevant for further discussion. What is relevant, is that, given an arbitrary uppertriangular matrix $\Delta$ with zero diagonal, we can easily reconstruct $\matr E$.

Our idea now is to use transformations  \eqref{eq:L_transformations} for the reduction of the matrix $\matr{L}$ to some suitable form, allowing for simple determination of the appropriate projector $\matr{P}$ for the rank-reducing transformation \eqref{eq:balance}. Namely we have the following
\begin{claim}\label{th:L0_canonical_form}
Using the transformations  \eqref{eq:L_transformations} it is possible to secure that $(\matr{L}_0)_{jk}=0$ for any $j$ and $k$ satisfying \begin{equation}
j\not\in S \& k\in S\cup\{k_0\}\,,
\end{equation}
where $k_0$ is a number of nontrivial Jordan cell (so that $n_{k_0}\neq0$) and $S$ is some set of the numbers of trivial Jordan cells, i.e. for any $i\in S$ holds $n_i=0$.
\end{claim}

A constructive proof of this claim is given in Algorithm \ref{alg:L0_to_Normal}.

\begin{algorithm}[H]
\caption{Reducing $\matr{L}_0$}
\label{alg:L0_to_Normal}
\SetKwInOut{Input}{Input}\SetKwInOut{Output}{Output}
\Input{Matrix $\matr{L}_0$ and integer $r$, such that  $\matr{L}_1=\diag(\underbrace{0,\ldots,0}\limits_{r},1,\ldots ,1)$ and  \eqref{eq:BF_condition} holds.}

\Output{$\{k_0,S,\Delta \}$, where ${\Delta}$ is uppertriangular  with zero diagonal such that the transformation \eqref{eq:L_general_transformation} results to $L_0$ of the form described in Claim~\ref{th:L0_canonical_form} with the corresponding $k_0$ and $S$.
}
\Begin {
	$S\longleftarrow \emptyset$\\
	$\Delta \longleftarrow $ zero matrix.\\
	\Repeat{$i\leqslant r$}{
		Construct $\widetilde{\matr L}_0=(a_1,a_2,\ldots)$ by striking out from $\matr{L}_0$ all rows with numbers from $S$. Below $a_i$ denotes the $i$-th column of this matrix.\\
		Find the minimal $i$ such that $i\not\in S$ and $i$-th column of $\widetilde{\matr L}_0$ is linearly dependent on first $i-1$ columns: $ a_i=c_1 a_1+\ldots+c_{i-1}a_{i-1} $.\\
		$\Delta_0\longleftarrow-c_1 \Delta^{(1,i)}-\ldots-c_{i-1}\Delta^{(i-1,i)}$\\
		$\widetilde\Delta_0\longleftarrow-c_1 \delta_{n_1n_i}\Delta^{(1,i)}-\ldots-c_{i-1}  \delta_{n_{i-1}n_i}\Delta^{(i-1,i)}$\\
		$\matr{L}_0\longleftarrow (\matr{I}-\widetilde\Delta_0)\matr{L}_0(\matr{I}+\Delta_0)$\\
		$\Delta \longleftarrow \Delta+\Delta_0+\Delta\Delta_0$\\
		$S\longleftarrow S\cup\{i\}$
	}
	\Return{ 
	$\{i,S\slash\{i\},\Delta \}$
	}
	
}
\end{algorithm}
The transformation on line 9 guarantees that any $i$-th column of $\widetilde{L}_0$  with $i\in S$ is zero. It may be not obvious why it is always possible to find appropriate $i$ on line 6 when $S$ contains only numbers larger than $r$. To explain this, let us examine the form of the matrix $\matr{L}(\lambda)$ after $m$ passes of the `repeat' loop. Then $S=\{i_1,\ldots i_m\}$, where $i_j>r$ is the number appearing at pass $\#j$. Let $\matr{L}^\prime(\lambda)$ denote a matrix obtained from $\matr{L}(\lambda)$ by simultaneous rearrangement of columns and rows in such a way that $i_k$-th column (and row) of the latter is $k$-th-to-last of the former. Then $\matr{L}^\prime(\lambda)$ has the following block form
\begin{equation}
\matr{L}^\prime(\lambda)=
\left(
\begin{array}{c|c}
\matr{X}(\lambda) & 
0 \\
 \hline
\matr Y & \matr Z(\lambda)
\end{array}
\right)\,,
\end{equation}
where $Z(\lambda)$ is a lower-triangular $m\times m$ matrix with diagonal elements equal to $\lambda$. Then, from the condition $\det \matr{L}^\prime(\lambda)=\det \matr{L}(\lambda)=0$, we obtain $\det \matr{X}(\lambda)=0$, and, in particular, 
\begin{equation}\label{eq:xcond}
\det \matr{X}(0)=0\,.
\end{equation}
Now we note that the columns of $\matr{X}(0)$ coincide, up to rearrangement, with the eligible columns of $\widetilde{\matr L}_0$ on line 5 of the algorithm, and the condition \eqref{eq:xcond} tells that there is a linear dependency between them. Thus, it is indeed possible to find $i$ as prescribed in line 6. The algorithm terminates at most when all $i>r$ are already included in $S$.\qed

Now we can use the output of Algorithm \ref{alg:L0_to_Normal} for the construction of the appropriate projector, such that the transformation \eqref{eq:balance} strictly lowers the rank of $\matr{A}_0$. First,  we use $\Delta$ for the reconstruction of the matrix $\matr E$. To this end it suffices to represent $\Delta$ as a linear combination of ${\Delta}^{(l,k)}$. Trivially, $\Delta=\sum_{l,k;\,l<k}\Delta_{lk}{\Delta}^{(l,k)}$, so $\matr E=\sum_{l,k;\,l<k}\Delta_{lk} \matr{E}^{(l,k)}$.
Using this matrix, we apply transformation \eqref{eq:U_general_transformation} to $\matr U$. Let now $u_k^{(\alpha)}$ and $v_k^{(\alpha)}$ be defined via Eqs. \eqref{eq:u_viaU} and \eqref{eq:v_viaU} for the transformed $\matr{U}$.
\begin{claim}\label{th:rank_reducing}
The transformation generated by
\begin{equation}\label{eq:T_Poincare_rank}
\matr T = \matr{\bal} (\matr P,0,x_2|x)\,,
\end{equation} 
where $x_2\neq0$ and
\begin{equation}\label{eq:P_Poincare_rank}
\matr P=\sum_{k\in S\cup\{k_0\}}u_{k}^{(0)}v_{k}^{(n_k)\dagger}
=u_{k_0}^{(0)}v_{k_0}^{(n_{k_0})\dagger}+\sum_{k\in S}u_{k}^{(0)}v_{k}^{(0)\dagger}
\,
\end{equation}
strictly lowers the rank of $\matr A_0$.
\end{claim}
The proof is very simple. We note that $\matr A_0 \matr P =0$ and the Laurent expansion of the transformed matrix $\widetilde{\matr{M}}$ near $x=0$ has the form
\begin{equation}
\widetilde{\matr{M}}(x)=\widetilde{\matr{A}}_0x^{-p-1}+O(x^{-p})\,,
\end{equation}
where
\begin{equation}
\widetilde{\matr A}_{0}=\overline{\matr P}\matr A_{0}+\overline{\matr P}\matr A_{1}\matr P\,.
\end{equation}
In order to prove that $\widetilde{\matr A}_{0}$ has matrix rank strictly smaller than that of $\matr A_{0}$ it is sufficient to demonstrate that $\widetilde{\matr A}_{0}$ has more eigenvectors (with zero eigenvalue) than $\matr A_{0}$. Let us check that any left eigenvector $v_j^{(0)\dagger}$ of $\matr A_{0}$  remains an eigenvector of $\widetilde{\matr A}_{0}$. This is obvious for $j\in S$ since $v_{j\in S}^{(0)\dagger}\overline{\matr{P}}=0$. Let now $j\not\in S$. Then $v_{j}^{(0)\dagger}\overline{\matr{P}}=v_{j}^{(0)\dagger}$ (in particular, this is valid for $j=k_0$ since $v_{k_0}^{(0)\dagger}u_{k_0}^{(0)}=0$). Then
\begin{equation}
v_j^{(0)\dagger}\widetilde{\matr A}_{0}=
v_j^{(0)\dagger}(\matr A_{0}+\matr A_{1}\matr P)=v_j^{(0)\dagger}\matr A_{1}\matr P=\sum_{k\in S\cup\{k_0\}}(\matr L_0)_{jk}v_{k}^{(0)\dagger}\qquad(j\not\in S)
\end{equation}
But, according to the Claim \ref{th:L0_canonical_form}, $(\matr L_0)_{jk}=0$ in the sum. So, we have proved that all eigenvectors of  $\matr A_0$ remain to be the eigenvectors of $\widetilde{\matr A}_0$. Obviously, we have an extra eigenvector of the latter, namely, $v_{k_0}^{(n_{k_0})\dagger}$, since $v_{k_0}^{(n_{k_0})\dagger}\overline{\matr{P}}=0$.\qed

Applying \eqref{eq:T_Poincare_rank} several times, we lower the rank of the leading coefficient $\matr A_0$ until it becomes zero (and thus  $\matr A_0$ itself is zero). This lowers the Poincar\'e rank by one. Acting in the same way, we finally lower the Poincar\'e rank to zero.

Algorithm \ref{alg:L0_to_Normal} as well as the transformation \eqref{eq:T_Poincare_rank} are very similar to those presented in Refs. \cite{BarkatouPfluegel2007,BarkatouPfluegel2009}. Moreover, our transformation is not optimal in a sense of \cite{BarkatouPfluegel2007}. The only advantage of our transformation  \eqref{eq:T_Poincare_rank} is that it gives as few terms in the sum in Eq. \eqref{eq:P_Poincare_rank} as possible. This will be helpful for the constructions of Section \ref{sec:global}.

\subsection*{Normalizing eigenvalues in Fuchsian singularities}

The results of the previous subsection allow one to reduce the Poincar\'e rank at one point in a stepwise manner provided $\matr{A}_0$ is nilpotent and  \eqref{eq:crit1} holds. If at some step either of these two conditions fails, then the point is irregular. Otherwise, we can lower Poincar\'e rank to zero, i.e., make system Fuchsian at a given point. The question remains whether we can do still better --- can we find a rational transformation that will restrict the form of the matrix residue? In this subsection we assume that $p=0$ in Eq. \eqref{eq:Mseries0}, i.e., that the Laurent series expansion of $\matr{M}(x)$ near $x=0$ has the form
\begin{equation}\label{eq:Mseries0a}
\matr{M}(x)=\matr{A}_0/x+\matr{A}_1+O(x)\,.
\end{equation}
Similar to the previous subsection, let 
\begin{equation}
\label{eq:basis}
\{u_k^{(\alpha)}|k=1\ldots N,\alpha=0,\ldots n_k\}
\end{equation}be a basis constructed of the generalized eigenvectors of $\matr{A}_0$ with the properties
\begin{equation}
\matr{A}_0 u_k^{(0)}=\lambda_k u_k^{(0)}\,,\quad \matr{A}_0 u_k^{(\alpha+1)}=\lambda_k u_k^{(\alpha+1)}+u_k^{(\alpha)}.
\label{eq:uproperties1}
\end{equation}
The vectors of the dual basis $\{v_1^{(n_1)},\ldots ,v_1^{(0)},v_2^{(n_2)},\ldots,v_2^{(0)},\ldots\}$ obey orthonormality condition \eqref{eq:uvorthonormality} and satisfy
\begin{equation}
v_k^{(0)\dagger}\matr{A}_0 =\lambda_k v_k^{(0)\dagger}\,,\quad v_k^{(\alpha+1)\dagger}\matr{A}_0 =\lambda_k v_k^{(\alpha+1)\dagger}+v_k^{(\alpha)\dagger}.
\label{eq:vproperties1}
\end{equation} 

Let us consider the transformation generated by $\matr{\bal}(\matr{P},0,x_2|x)$, where 
\begin{equation}\label{eq:p}
\matr P=u_1^{(0)}v_1^{(n_1)\dagger}\,.
\end{equation} 
Since $\overline{\matr{P}}\matr{A}_0 \matr{P}= \lambda_1\overline{\matr{P}}\matr{P} = 0$, the Laurent series expansion near $x=0$ of the transformed matrix $\widetilde{\matr{M}}$ starts from $x^{-1}$:
\begin{equation}\label{eq:Mseries0t}
\widetilde{\matr{M}}(x)=\widetilde{\matr{A}}_0/x+O(x^{0})
\end{equation}
with
\begin{equation}\label{eq:A_p1}
\widetilde{\matr{A}}_0=\overline{\matr{P}}\matr A_{0}+\matr A_{0}\matr{P} +\matr{P} +\overline{\matr{P}}\matr A_{1}\matr{P}
\end{equation}
\begin{prop}\label{th:prop}
With the account of multiplicity, only one eigenvalue of $\widetilde{\matr{A}}_0$ is different from the corresponding eigenvalue of $\matr{A}_0$. Namely, $\lambda_1$ changes to $\lambda_1+1$.
\end{prop}
The proof of this proposition becomes obvious if one examines the form of $\widetilde{\matr{A}}_0$ in the basis \eqref{eq:basis} and calculates its characteristic polynomial. Indeed, in the basis \eqref{eq:basis}, matrix $\matr{A}_0$ has the following form $\matr{A}_0=\diag(\lambda_1,\ldots )+\diag^{(1)}(f_1,f_2,\ldots )$, where $\diag^{(1)}$ denotes the matrix with $f_1,f_2,\ldots$ standing above the diagonal and zero elsewhere, $f_i=0 \text{ or } 1$. Then
\begin{equation}
\widetilde{\matr{A}}_0=c_1\otimes(1,0,\ldots)+\diag(\lambda_1+1,\ldots )+\diag\nolimits^{(1)}(0,f_2,\ldots )\,,
\end{equation}
where $c_1$ is the first column of the matrix $\matr{A}_1$. So, the matrix $\widetilde{\matr{A}}_0$ differs from $\matr{A}_0$ only in the first column and first row. Obviously, the characteristic polynomial of the former is $P(\widetilde{\matr{A}}_0,\lambda)=(\lambda_1+1-\lambda)P(\matr{A}_0,\lambda)/(\lambda_1-\lambda)$.\qed

Similar, $\matr{\bal}(u_1^{(n_1)}v_1^{(0)\dagger},x_2,0|x)$ shifts one eigenvalue down. Thus we come to the following
\begin{claim}
Using balances
\begin{gather}
\matr{\bal}(u_1^{(0)}v_1^{(n_1)\dagger},0,x_2|x)\,,\nonumber\\
\matr{\bal}(u_1^{(n_1)}v_1^{(0)\dagger},x_2,0|x)\,,\label{eq:eigenvalue_shifting}
\end{gather} 
it is possible to reduce the matrix residue to the normalized form in which all its eigenvalues have the real parts lying in the interval $[a,a+1)$, where $a$ is a real number.
\end{claim}
Usual choice is $a=0$, however we will prefer $a=-1/2$ due to the reasons which should be clear from the consideration below.
Note that in this normalized form the monodromy matrix for the small loop around $x=0$ is given, up to similarity, by 
\begin{equation}\label{eq:monodromy}
\mathcal{M}=\exp[2\pi i\matr{A}_0]
\end{equation}

Thus, using the results of this subsection and the previous one, we can simply find the monodromy matrix around any regular point of the differential system. In particular, we can detect whether a given point is an apparent singularity (i.e., the monodromy is an identity). To this end, we note that, given $\matr{A}_0$ is normalized and Eq. \eqref{eq:monodromy} defines an identity matrix, one may easily conclude that $\matr{A}_0=0$ (by considering the matrix function of the Jordan form). Therefore, normalization totally eliminates any apparent singularity. Note that if the matrix residue is not normalized, in general, the monodromy matrix is not given by Eq. \eqref{eq:monodromy} due to resonances (the eigenvalues of $\matr{A}_0$, whose difference is an integer number).

\section{Global reduction}\label{sec:global}

The transformations considered in the previous section have a serious flaw: while improving the form of the matrix at one point, they, in general, worsen its form in another. In principle, the reduction of the Poincar\'e rank to zero can always be done at the cost of introducing some apparent Fuchsian singularities. This is because balances may increase the pole order at most by one. So, choosing at each step a regular point as $x_2$, we can globally reduce the Poincar\'e rank to zero. However, we, of course, would like to avoid generating unnecessary apparent singularities in the process of reducing the Poincar\'e rank. The situation is different when we want to normalize all Fuchsian singularities. In this case we definitely do not want to generate apparent singularities, since any apparent singularity is not normalized (otherwise there would be no singularity at all). In the present section we show that, except for some degenerate cases, it is possible to slightly modify the projectors constructed in the previous section so that the resulting balances respect the Poincar\'e rank at the second point.

Let us first describe transformations which do not increase  Poincar\'e rank at any point. Suppose $x_1$ and $x_2$ are two finite singular points of the matrix $\matr{M}(x)$, so that the Laurent series around $x_1$ and $x_2$ have the form
\begin{align}
\matr{M}(x)&=\matr{A}_0(x-x_1)^{-p_1-1}+O((x-x_1)^{-p_1})\\
\matr{M}(x)&=\matr{B}_0(x-x_2)^{-p_2-1}+O((x-x_2)^{-p_2})\label{eq:Mseries2}
\end{align}
and $ p_1\geqslant0\,,\  p_2\geqslant0\,$. 
\begin{claim}\label{th:conserving_condition}
If $\matr{Q}$ is a projector such that $\mathop\mathrm{Im}\matr{Q}$ and $\mathop\mathrm{Ker}\matr{Q}$ are invariant subspaces of $\matr{A}_0$ and $\matr{B}_0$, respectively, then the transformation $\matr{\bal}(\matr{Q},x_1,x_2|x)$ does not increase the Poincar\'e rank of $\matr{M}$ at any point.
\end{claim}
The proof is straightforward after observing that $\matr{Q}$ satisfies
\begin{equation}
\overline{\matr{Q}}\matr{A}_0\matr{Q}=\matr{Q}\matr{B}_0\overline{\matr{Q}}=0\,.\label{eq:Pcond1}
\end{equation}
We stress that the claim is also valid when one or both points are Fuchsian.

More explicitly, let $\{u_1,\ldots, u_m\}$ span $m$-dimensional invariant space of $\matr{A}_0$. Suppose that, among $m$-dimensional left invariant spaces of $\matr{B}_0$, there is one which allows for the basis $\{v_1^\dagger,\ldots, v_m^\dagger\}$ satisfying
\begin{equation}
v_j^\dagger u_k=\delta_{jk}
\end{equation}
Such a basis for $m$-dimensional left space exists iff the space does not contain a vector, orthogonal to all $u_1,\ldots, u_m$.
Then 
\begin{equation}
\matr{Q}=\sum_{k=1}^m u_k v_k^\dagger
\end{equation}
is the projector satisfying conditions of Claim \ref{th:conserving_condition}.

Let us now consider the $\matr{Q}$-balance between $x_1$ and $x_2$ with 
\begin{equation}\label{eq:P0}
\matr{Q}=\sum_{k\in S\cup\{k_0\}}u_{k}^{(0)}v_k^\dagger\,
\end{equation}
where all notations are as in Eq. \eqref{eq:P_Poincare_rank} except that now $v_k^\dagger$ span some left-invariant space of $\matr{B}_0$, but still satisfy $v_j^\dagger u_k^{(0)}=\delta_{jk}$. 
\begin{claim}\label{th:rank_reducing_modified}
Let $\matr{M}(x)$ has Laurent series expansion near $x=0$ as in \eqref{eq:Mseries0} with $p>0$ and that near $x=x_2$ as in \eqref{eq:Mseries2}. Then the $\matr Q$-balance between $0$ and  $x_2$, Eq. \eqref{eq:balance} with $\matr Q$ from Eq. \eqref{eq:P0}
strictly diminishes the matrix rank of $\matr A_0$ and does not increase the Poincar\'e rank at any other point.
\end{claim}
In order to prove this claim, let us use the identities
\begin{equation}\label{eq:PQ_identities}
\matr P\matr Q=\matr Q\,,\quad \matr Q\matr P=\matr P
\end{equation}
and
\begin{equation}
\matr A_0\matr Q=\matr A_0\matr P=0\,.
\end{equation}
These identities simply follow from the definitions of the projectors $\matr P$ and $\matr Q$, Eqs. \eqref{eq:P_Poincare_rank} and \eqref{eq:P0}. Then
\begin{multline}
\widetilde{\matr A}_0=\overline{\matr Q}\matr A_{0} +\overline{\matr Q}\matr A_{1}\matr Q 
=(\overline{\matr Q}+\matr P)\overline{\matr P}\matr A_{0} +(\overline{\matr Q}+\matr P)\overline{\matr P}\matr A_{1}\matr P (\overline{\matr P}+\matr Q)
=(\overline{\matr Q}+\matr P)[\overline{\matr P}\matr A_{0}+\overline{\matr P}\matr A_{1}\matr P](\overline{\matr P}+\matr Q)
\end{multline}
The expression in square brackets is just the transformation of the leading coefficient generated by $\matr{\bal}(\matr{P},0,x_2|x)$. Taking into account that $(\overline{\matr Q}+\matr P)=(\overline{\matr P}+\matr Q)^{-1}$, we see that the transformed leading coefficient $\widetilde{\matr A}_0$ after the transformation $\matr T_1=\matr{\bal}(\matr{Q},0,x_2|x)$ coincides with that after the transformation $\matr T_2=\matr{\bal}(\matr{P},0,x_2|x)(\overline{\matr P}+\matr Q)$ (Note that these transformations are nevertheless different, since $\matr T_1=(\overline{\matr Q}+\matr P)\matr T_2$). Then, the correctness of Claim \ref{th:rank_reducing_modified} follows from that, on one hand, $\matr{\bal}(\matr{Q},0,x_2|x)$ satisfies conditions of Claim \ref{th:conserving_condition}, and on the other hand the leading coefficient is transformed as though by the transformation which is a product of $\matr{\bal}(\matr{P},0,x_2|x)$, satisfying conditions of Claim \ref{th:rank_reducing}, and constant nonsingular matrix (which does not change the rank of $A_0$).\qed

Similar modifications should also be made for the balances \eqref{eq:eigenvalue_shifting} used for the normalization of the matrix residue eigenvalues. We simply replace in their definitions the vectors $v_1^{(n_1)\dagger}$ and $u_1^{(n_1)}$ with $v^{\dagger}$ and $u$ which are left and right eigenvectors of the matrix $\matr{B}_0$, respectively, provided they satisfy $v^{\dagger}u_1^{(0)}=1$ and $v_1^{(0)\dagger}u=1$. 
\begin{claim}\label{th:normalizing_modified}
Let $\matr{M}(x)$ has Laurent expansion near $x=0$ as in \eqref{eq:Mseries0a} and that near $x=x_2$ as in \eqref{eq:Mseries2}. Let $u$ and $v^{\dagger}$ be the right and left eigenvectors of $A_0$ and $B_0$, respectively. Then the $\matr{\bal}(uv^{\dagger},0,x_2|x)$ increases by one the eigenvalue of $\matr A_0$, corresponding to $u$, and does not increase the Poincar\'e rank at any point.
\end{claim}
The proof is very similar to the previous case. Let now $\matr Q=uv^{\dagger}$ and $\matr P$ be defined in \eqref{eq:p} with $u_1^{(0)}=u$. In addition to the identities \eqref{eq:PQ_identities} we use now
\begin{equation}
\matr A_0\matr Q=\lambda\matr Q\,,\quad
\matr A_0\matr P=\lambda \matr P\,.
\end{equation}
Then
\begin{multline}
\widetilde{\matr A}_0=\overline{\matr Q}\matr A_{0} 
+\underbrace{\matr A_{0}\matr Q+\matr Q}_{\propto \matr Q=(\overline{\matr Q}+\matr P)\matr Q} 
+\overline{\matr Q}\matr A_{1}\matr Q 
=(\overline{\matr Q}+\matr P)\overline{\matr P}\matr A_{0} +(\overline{\matr Q}+\matr P)(\matr A_{0}+\matr I)\matr Q +(\overline{\matr Q}+\matr P)\overline{\matr P}\matr A_{1}\matr P
(\overline{\matr P}+\matr Q)\\
=(\overline{\matr Q}+\matr P)\overline{\matr P}\matr A_{0} +(\overline{\matr Q}+\matr P)(\matr A_{0}+\matr I)\matr P(\overline{\matr P}+\matr Q) +(\overline{\matr Q}+\matr P)\overline{\matr P}\matr A_{1}\matr P
(\overline{\matr P}+\matr Q)\\
=(\overline{\matr Q}+\matr P)[\overline{\matr P}\matr A_{0} +\matr A_{0}\matr P+\matr P +\overline{\matr P}\matr A_{1}\matr P]
(\overline{\matr P}+\matr Q)\,,
\end{multline}
where in the last transition we used the identity $\overline{\matr P} \matr A_0=\overline{\matr P} \matr A_0(\overline{\matr P}+\matr Q)$.
Again, we see that the expression in square brackets is just the transformation of the leading coefficient generated by $\matr{\bal}(\matr{P},0,x_2|x)$. Since $\widetilde{\matr A}_0$ is, up to a similarity, the same as in \eqref{eq:A_p1}, the Proposition \ref{th:prop} proves the claim.\qed

If the second point is also Fuchsian, this transformation simultaneously shifts in the opposite direction the eigenvalue of the matrix $\matr{B}_0$, corresponding to $v^{\dagger}$ and $u$, respectively. Therefore, the process of normalization resembles balancing the scales, this is the reason why we call the transformation \eqref{eq:balance} the balance.
\begin{defi}\label{def:balanced_with}
We say that the Fuchsian point $x_1$\textbf{ can be balanced} with the singular point $x_2\neq x_1$ if at least one of the two conditions holds 
\begin{enumerate}
\item there exist $u$ and $v^\dagger$, right and left eigenvectors of $\matr A_0$ and $\matr B_0$, such that $v^{\dagger}u=1$ and the real part of the eigenvalue of $\matr A_0$, corresponding to $u$ is less than $-1/2$.
\item there exist $u$ and $v^\dagger$, right and left eigenvectors of $\matr B_0$ and $\matr A_0$, such that $v^{\dagger}u=1$ and the real part of the eigenvalue of $\matr A_0$, corresponding to $v^{\dagger}$ is greater or equal than $1/2$.
\end{enumerate}
Here $\matr A_0$ and $\matr B_0$ are the matrix residues of the Laurent expansion of $\matr M(x)$ near $x=x_1$ and $x=x_2$, respectively. More specific, we say  $x_1$ can be balanced with $x_2$ via $\matr{\bal}(uv^\dagger,x_1,x_2|x)$ or via $\matr{\bal}(uv^\dagger,x_2,x_1|x)$, depending on whether the first or second condition holds.
\end{defi}

\begin{defi}\label{def:mutually_balanced}
We say that two Fuchsian points $x_1$ and  $x_2\neq x_1$\textbf{ can be mutually balanced} if at least one of the two conditions holds 
\begin{enumerate}
\item there exist $u$ and $v^\dagger$, $\matr A_0 u=\lambda u$, $v^{\dagger}\matr B_0=\mu v^{\dagger}$, such that $\Re\lambda<1/2$, $\Re\mu\geqslant 1/2$, and $v^{\dagger}u=1$.
\item there exist $u$ and $v^\dagger$, $\matr B_0 u=\lambda u$, $v^{\dagger}\matr A_0=\mu v^{\dagger}$, such that $\Re\lambda<1/2$, $\Re\mu\geqslant 1/2$, and $v^{\dagger}u=1$.
\end{enumerate}
Here $\matr A_0$ and $\matr B_0$ are the matrix residues of the Laurent expansion of $\matr M(x)$ near $x=x_1$ and $x=x_2$, respectively. More specific, we say  that $x_1$ and $x_2$ can be mutually balanced via $\matr{\bal}(uv^\dagger,x_1,x_2|x)$ or via $\matr{\bal}(uv^\dagger,x_2,x_1|x)$, depending on whether the first or second condition holds.
\end{defi}

The reason for these definitions is clear: if $x_1$ can be balanced with some point, there exists a balance which moves one eigenvalue of matrix residue in $x=x_1$ towards the interval $[-1/2,0)$. If the two points can be mutually balanced, there exists a balance which moves one eigenvalue of matrix residue at $x=x_1$ and that at $x=x_2$ towards the interval $[-1/2,1/2)$.

\section{Reduction process}

The transformations described in two previous sections give one much freedom in reducing a given system to a Fuchsian form and in normalizing eigenvalues of the matrix residues at Fuchsian points.
Let us summarize the basic line of the reduction process in the form of two algorithms.

\begin{algorithm}[H]
\caption{Reduction to Fuchsian form}
\label{alg:Fuchsian}
\SetKwInOut{Input}{Input}\SetKwInOut{Output}{Output}
\Input{Matrix $\matr M(x)$ appearing in the right-hand side of the differential equation.}

\Output{Transformation matrix $\matr T(x)$ transforming  $\matr M(x)$ to $\widetilde{\matr{M}}(x)$, such that $\widetilde{\matr{M}}(x)$ is Fuchsian at any point.}
\Begin {
	$\widetilde{\matr{M}}\longleftarrow \matr M(x)$\\
	$\matr{T}\longleftarrow$ identity matrix\\
	\While{there is a point with positive Poincar\'e rank}
	{
		\If{there is a pair of singular points $x_1$ and $x_2$, such that 
		\begin{enumerate}
		\item Poincar\'e rank of the system at $x=x_1$ is positive
		\item It is possible to construct the projector $\matr Q$ as in Eq. \eqref{eq:P0}
		\end{enumerate}
		}{
			$\matr T_0\longleftarrow\matr{\bal}(\matr Q, x_1,x_2|x)$\\
			$\widetilde{\matr{M}}\longleftarrow\matr{T}_0^{-1}\widetilde{\matr{M}}\matr{T}_0-\matr{T}_0^{-1}\partial_{x}\matr{T}_0$\\
			$\matr T\longleftarrow\matr T\matr T_0$
		}
		\Else{
			Let  $x_1$ be the point with positive Poincar\'e rank.\\
			Choose arbitrary regular point $x_2$.\\
			$\matr T_0\longleftarrow\matr{\bal}(\matr P, x_1,x_2|x)$, where $\matr P$ is defined in Eq. \eqref{eq:P_Poincare_rank}\\
			$\widetilde{\matr{M}}\longleftarrow\matr{T}_0^{-1}\widetilde{\matr{M}}\matr{T}_0-\matr{T}_0^{-1}\partial_{x}\matr{T}_0$\\
			$\matr T\longleftarrow\matr T\matr T_0$
		}
	}
	\Return{$\matr T$}
}
\end{algorithm} 
Note that this algorithm assumes that all singular points of the system are regular, so the transformation on line 13 can be always constructed. Let us comment on the condition 2 on line 5. This condition holds if it is possible to find an invariant subspace of the matrix $\matr{B}_0$, which has a dual basis with $\{u_{k}^{(0)}, k\in S\cup \{l\}\}$, see \eqref{eq:P0}. It appears to be a nontrivial task due to the complexity of the set of invariant spaces of an arbitrary matrix, see, e.g. Ref. \cite{GohbergLancasterRodman1986}. However, one might try the subspace formed by the eigenvectors of $\matr B_0$, and consecutively add vectors from the Jordan chain if needed. If these attempts fail, one may simply go to line 10 with a penalty of possibly introducing an extra apparent singularity. Given that at the next stage this singularity is likely to disappear, this is not a real problem. 

Next stage is described by the following algorithm

\begin{algorithm}[H]
\caption{Normalization}
\label{alg:Normal}
\SetKwInOut{Input}{Input}\SetKwInOut{Output}{Output}
\Input{Matrix $\matr M(x)$ appearing in the right-hand side of the differential equation, having zero Poincar\'e rank at all singular points.}

\Output{Transformation matrix $\matr T(x)$ transforming  $\matr M(x)$ to $\widetilde{\matr{M}}(x)$, such that $\widetilde{\matr{M}}(x)$ is normalized at as many points as possible.}
\Begin {
	$\widetilde{\matr{M}}\longleftarrow \matr M(x)$\\
	$\matr{T}\longleftarrow$ identity matrix\\
	Detect apparent singularities using the transformations \eqref{eq:eigenvalue_shifting}\\
	Select a singular point $x_0$ which is not an apparent singularity. If there are only apparent singularities, let $x_0$ be one of them.\\
	\While{there is a pair of points which can be mutually balanced or there is a point which can be balanced with $x_0$}
	{
		\If{there is a pair of singular points $x_1$ and $x_2$, which can be mutually balanced
		}{
			Let $x_1$ and $x_2$ can be mutually balanced via $\matr T_0$.\\
			$\widetilde{\matr{M}}\longleftarrow\matr{T}_0^{-1}\widetilde{\matr{M}}\matr{T}_0-\matr{T}_0^{-1}\partial_{x}\matr{T}_0$\\
			$\matr T\longleftarrow\matr T\matr T_0$			
		}
		\Else{
			Let $x_1$ can be balanced with $x_0$ via $\matr T_0$.\\
			$\widetilde{\matr{M}}\longleftarrow\matr{T}_0^{-1}\widetilde{\matr{M}}\matr{T}_0-\matr{T}_0^{-1}\partial_{x}\matr{T}_0$\\
			$\matr T\longleftarrow\matr T\matr T_0$
		}
	}
	\Return{$\matr T$}
}
\end{algorithm} 

Though being very useful, the above algorithm does not necessarily give a canonical form of $\matr M(x)$ in any sense. In particular, the outcome depends on the sequence of the pairs of points chosen at a specific step. However, in many tested cases, this algorithm succeeds in normalizing the system at all but one singular points, in particular, removing all apparent singularities. As it was already mentioned, the possibility of removing all apparent points is equivalent to the content of the 21st Hilbert problem. As proved by Bolibrukh \cite{Bolibrukh1989}, this task is not always possible to complete and, therefore, the 21st Hilbert problem has a negative solution. In his paper Bolibrukh presents an example of the system which can not be reduced to Fuchsian form without apparent singularities. We have checked, that our algorithm indeed fails to reduce this system. At some step it appears to be not possible to balance an apparent singularity with any other singular point due to the orthogonality of the corresponding eigenvectors. 

On the other hand in the same paper it was proved that for $n=2$ the 21st Hilbert problem can always be solved. For our setup, it translates to the statement that, given a Fuchsian system of two equations, it is always possible to get rid of the apparent singularities. Let us show that the tools developed in this section easily allow one to perform this task, thus, giving a constructive proof of the statement. Our line of reasoning is very simple: we show that it is always possible to shift the eigenvalues of the matrix residue in the apparent singularity towards the interval $[-1/2,1/2)$ without introducing new apparent points and increasing the pole order. The eigenvalues of the matrix residue in apparent singularity should definitely be integer, otherwise, we may show that the point is not an apparent singularity by normalizing the system at this point (possibly spoiling its form in others) and calculating the monodromy from Eq. \eqref{eq:monodromy}. Moreover, when both eigenvalues are zero, the whole matrix should be zero. Then, in a finite sequence of shifts we will eventually eliminate singularity. Eliminating singularities one by one, we obtain the desired form. 

Suppose $x=0$ is the apparent singularity and $A_0\neq 0$ is a $2\times 2$ matrix residue at this point.  Note that the differential system in Fuchsian form can not have only one singular point, so we may rely on the existence of at least one singularity different from $x=0$. If both eigenvalues of $A_0$ are nonzero and of the same sign, we may use the transformation $\matr T=\frac{x}{x-x_2}\matr I$ or $\matr T=\frac{x-x_2}{x}\matr I$ to raise or lower both eigenvalues. Here $x_2$ is some other singular point. Thus, we may restrict ourselves to the case when, say, one eigenvalue is negative and the other one is non-negative. Suppose that $\matr A_0=\diag(n_1<0,n_2\geqslant 0)$. The right eigenvector of $\matr A_0$, corresponding to $n_1$ is $u=(1,0)^\dagger$. Suppose, all left eigenvectors of matrix residues at other singular points are orthogonal to $u$. Then, it is easy to show that the general form of these matrix residues is $\left(\begin{array}{cc}
a&b\\
0&a
\end{array}\right)$. 
But this form is in obvious contradiction with the requirement that the sum of all matrix residues is zero. This is because the diagonal elements of this sum are $n_1+\sum_i a_i$ and  $n_2+\sum_i a_i$ which can not be both zero. Therefore, there is a left eigenvector $v^\dagger$ of the matrix residue at some point $x_2$, such that $v^\dagger u=1$ and $x=0$ can be balanced with $x=x_2$ via $\matr{\bal}(u v^\dagger,0,x_2|x)$.

\section{Factoring out $\epsilon$}\label{sec:factoring_epsilon}

So far, we described the constructions which are not specific to the systems depending on parameter. However, the idea of their application to the reduction of the systems, depending on $\epsilon$, should be clear. First, we use Algorithm \ref{alg:Fuchsian} to reduce the system to Fuchsian form. A necessary condition of existence of the $\epsilon$-form \eqref{eq:CanonicalForm} is that the eigenvalues of all matrix residues have the form $n+\alpha \epsilon$, where $n$ is integer. If this condition is not satisfied, then the system definitely can not be transformed to the form \eqref{eq:CanonicalForm}. In this case one might try some changes of variable\footnote{Note that such a situation often happens for the integrals with massive internal lines. When passing back to the original variable one encounters transformations, involving algebraic functions (in particular, square roots).}. If the condition holds, one may pass to the Algorithm \ref{alg:Normal} in order to normalize eigenvalues of the matrix residue at all but one point $x=x_1$, assuming $\epsilon$ is sufficiently small (i.e., assuming $n+\alpha \epsilon$ belongs to the interval $[-1/2,1/2)$ only if $n=0$). If this step appears to be doable, the normalized eigenvalues are all proportional to $\epsilon$. The sum of the eigenvalues in $x=x_1$ is also proportional to $\epsilon$ since the matrix residue at this last point is simply minus the sum of the matrix residues at the normalized points (and so the trace is minus sum of the traces). Then one should try to balance $x=x_1$ in two steps. First, shift down one of the positive unnormalized eigenvalues by means of balance with some point $x=x_2$, either singular or regular, and then mutually balance $x_1$ and $x_2$ shifting up one of the negative unnormalized eigenvalues of the matrix residue at $x=x_1$. 

Let us assume from now on that it appeared to be possible to secure by the above method that the system is Fuchsian and normalized at all points. Then we have a system 
\begin{equation}\label{eq:Fuchsian}
\partial_{x}\mathbf{J}=\sum_{k}\frac{\matr{M}_k(\epsilon)}{x-x_k}\mathbf{J}\,,
\end{equation}
and the eigenvalues of all matrices $\matr{M}_k$ are proportional to $\epsilon$. Clearly, this does not necessarily mean that matrices $\matr{M}_k$ themselves are proportional to $\epsilon$. If we had only one matrix $\matr M_1(\epsilon)$, we could have factorized $\epsilon$ by making a transformation which transforms $\matr M_1(\epsilon)/\epsilon$ to Jordan form. In general case we need to find an $x$-independent transformation matrix which simultaneously transforms all matrices $\matr{M}_k(\epsilon)$ to the form $\epsilon\matr{S}_k$, where $\matr{S}_k$ are constant matrices\footnote{Note that any $x$-dependent rational transformation necessarily has at least one singular point and shifts the eigenvalues of the matrix residue in this point thus spoiling normalization. Normalization, in turn, necessarily holds for the $\epsilon$-form.}. Let $\matr T(\epsilon)$ be such a matrix. Then we have
\begin{equation}\label{eq:T_constant}
\matr T^{-1}(\epsilon)\frac{\matr{M}_k(\epsilon)}{\epsilon}\matr T(\epsilon)=\matr{S}_k=\matr T^{-1}(\mu)\frac{\matr{M}_k(\mu)}{\mu}\matr T(\mu)\,.
\end{equation}
Multiplying this equation by $\matr T(\epsilon)$ from the left and by $\matr T^{-1}(\mu)$ from the right, we obtain a linear system
\begin{gather}\label{eq:constant_transformation}
\frac{\matr{M}_1(\epsilon)}{\epsilon}\matr T(\epsilon,\mu)=\matr T(\epsilon,\mu)\frac{\matr{M}_1(\mu)}{\mu}\,,\nonumber\\
\vdots\nonumber\\
\frac{\matr{M}_m(\epsilon)}{\epsilon}\matr T(\epsilon,\mu)=\matr T(\epsilon,\mu)\frac{\matr{M}_m(\mu)}{\mu}
\end{gather}
for the elements of the matrix $\matr T(\epsilon,\mu)=\matr T(\epsilon)\matr T^{-1}(\mu)$. If the general solution of this system (found routinely) determines an invertible matrix, the transformation we are looking for can be chosen as $\matr T(\epsilon)=\matr T(\epsilon,\mu_0)$, where $\mu_0$ is some arbitrarily chosen number, provided $\matr T(\epsilon,\mu)$ is nonsingular at $\mu=\mu_0$.

\section{Using block-triangular form}\label{sec:btform}

The size $n$ of the matrices $\matr M(\epsilon,x)$ appearing in the differential equations for master integrals may be quite large ($\sim$ several tens). This may constitute computational complications for the transformations that we need. Fortunately, the very process of the derivation of the differential equations, the IBP reduction, shows that $\matr M(\epsilon,x)$ contains a lot of zeros. Namely, the integral $J_1$ may enter the right-hand side of the differential equation for the integral $J_2$ only if the graph corresponding to $J_1$ can be obtained from that corresponding to $J_2$ by contraction of some edges. In particular, this means that the matrix $\matr M(\epsilon,x)$ has a block-triangular form with diagonal blocks corresponding to the integrals with a given set of denominators (= integrals of a given sector). 

Let us show that we can use this block-triangular form to essentially alleviate the process of reduction. Suppose from now on that we have already reduced all diagonal blocks of $\matr M(\epsilon,x)$ to $\epsilon$-form.  Basically, the idea of further reduction is simple. In order to reduce the pole order of the off-diagonal elements we redefine the integrals by adding some suitable combination of the simpler integrals, similar to the approach of Refs. \cite{Caron-HuotHenn2014,GehrmannManteuffelTancrediWeihs2014}. Let us prove that it is always possible to make this redefinition in order to reduce the Poincar\'e rank at a given point to zero without changing both the block-triangular structure of the system and the Poincar\'e rank at other points. Therefore, it gives one a tool to reduce the system to Fuchsian form. 

We prove by the induction over sectors. Without generality loss, we may assume that we are interested in reducing the Poincar\'e rank to zero at $x=0$ \footnote{In what follows, when speaking about singularity and Poincar\'e rank we often omit references to $x=0$ for brevity.}. 
Suppose $\mathbf{J}_1$ is a column-vector of master-integrals in a certain sector $\boldsymbol \theta$. By the induction hypothesis the differential system for the integrals in the subsectors of  $\boldsymbol \theta$ already has zero Poincar\'e rank and thus no master in the subsectors will not be changed at this and later steps.
We can write the differential system for $\mathbf{J}_1$ in the form
\begin{equation}\label{eq:triangular}
x \partial_x\mathbf{J}_1=\epsilon \matr A(x) \mathbf{J}_1 + x^{-r}\matr B(\epsilon) \mathbf{J}_2+\ldots\,,
\end{equation}
where $\mathbf{J}_2$ is the column-vector of the master integrals in the most complex subsector of $\boldsymbol \theta$ entering the right-hand side of the equation with singular coefficient, whose Laurent expansion starts with $x^{-r}\matr B(\epsilon)$ with $r>0$. By the assumption, $\matr A(x)$ is regular at $x=0$. Naturally, the number of entries in $\mathbf{J}_1$ and  $\mathbf{J}_2$ is not required to be the same, so, in general, $\matr B$ is a rectangular matrix. In Eq. \eqref{eq:triangular} the dots denote terms which are either nonsingular, or contain integrals in the less complex sectors than the sector of $\mathbf{J}_2$, or contain integrals  $\mathbf{J}_2$ with coefficients less singular than $x^{-r}$. 
The differential equation for $\mathbf{J}_2$ has the form
\begin{equation}
x \partial_x\mathbf{J}_2=\epsilon \matr C(x) \mathbf{J}_2 +\ldots\,,
\end{equation}
where $\matr C(x)$ is regular at $x=0$. The dots denote contribution of the subsectors.
Let us make the substitution 
\begin{equation}\label{eq:subst}
\mathbf{J}_1=\widetilde{\mathbf{J}}_1+x^{-r}\matr D \mathbf{J}_2\,,
\end{equation} where $\matr D$ is a constant matrix. We have
\begin{equation}
x \partial_x\widetilde{\mathbf{J}}_1=\epsilon \matr A(x) \widetilde{\mathbf{J}}_1+x^{-r}\left[\matr B(\epsilon)+ r \matr D+\epsilon \matr A(x) \matr D -\epsilon  \matr D\matr C(x)\right]\mathbf{J}_2+\ldots\,.
\end{equation}
Therefore, in order to cancel $x^{-r}$ singularity, we need to find such $\matr D$ that
\begin{equation}
\matr D+\frac{\epsilon}{r} [\matr A(0) \matr D -\matr D\matr C(0)]=-\frac1r\matr B(\epsilon)
\end{equation} 
This is a system of linear equations for the matrix elements of $\matr D$. This system obviously has a solution since the linear operator acting on $\matr D$ in the right-hand side is arbitrarily close to unity. Note that this line of reasoning does not work when the diagonal blocks are not in $\epsilon$-form and/or when $r=0$. Therefore, starting from the most complex integrals in the right-hand side and from the highest poles in their coefficients, we can eliminate singular coefficients in the right-hand side, step-by-step. Note that the substitution \eqref{eq:subst} corresponds to the transformation generated by 
\begin{equation}\label{eq:Tn}
\matr T=\matr I+\frac{\matr N}{x^r}\,,
\end{equation}
where $\matr N$ is a matrix whose nonzero elements coincide with the elements of $\matr D$. It is easy to see that $\matr N^2=0$, so that the inverse matrix has the form
\begin{equation}
\matr T^{-1}=\matr I-\frac{\matr N}{x^r}\,.
\end{equation}
Therefore, this transformation is regular everywhere, except $x=0$.

Now we may assume that we have a Fuchsian block-triangular matrix $\matr M(\epsilon, x)$ such that each diagonal block is in $\epsilon$-form. Since the characteristic polynomial of this matrix is a product of those of the diagonal blocks, the eigenvalues of $\matr M(\epsilon, x)$ are proportional to $\epsilon$ and we have a system of the form \eqref{eq:Fuchsian}. In order to find a transformation matrix $\matr T(\epsilon)$ from \eqref{eq:T_constant}, which, in addition, preserves the block-triangular form of $M(x)$, we may nullify in all elements of $\matr T(\epsilon,\mu)$, corresponding to zero elements of $\matr M(\epsilon,x)$, before solving the system \eqref{eq:constant_transformation}.

\section{Example}\label{sec:example}
\begin{figure}[h]
\centering
\includegraphics{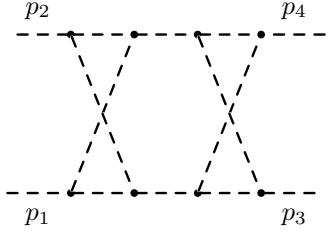}
\caption{Three-loop "XX-box" topology. Internal dashed lines denote massless propagators, $p_1^2=p_2^2=p_3^2=p_4^2=0$, $(p_1+p_2)^2=s$, $(p_1-p_3)^2=t$.}
\label{fig:top}
\end{figure}

\begin{figure}[h]
\includegraphics{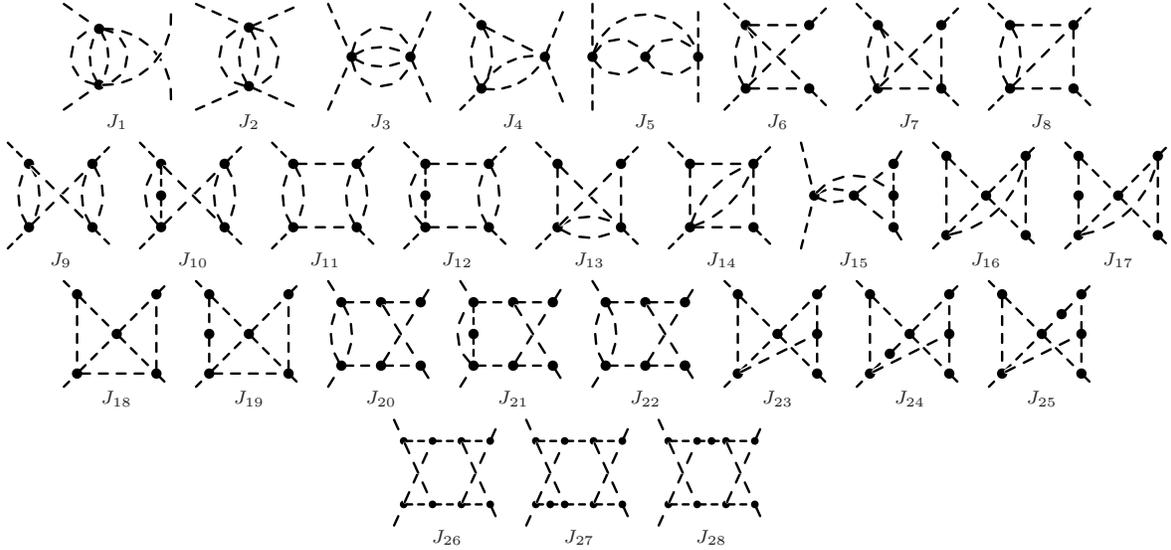}

\caption{Master integrals of the topology in Fig. \ref{fig:top}. Integrals $J_{26-28}$ are determined with the help of \texttt{Mint}.}
\label{fig:mis}
\end{figure}

Let us demonstrate in some details how our method works for the master integrals in the topology shown in Fig. \ref{fig:top}. There are 28 master integrals shown in Fig. \ref{fig:mis}.
We use an experimental version of \texttt{LiteRed}, \cite{Lee2013a,Lee2012}, for the IBP reduction.   Unfortunately, due to the complexity of the IBP reduction, we have not been able to obtain starting differential equations for the 3 master integrals in the highest sector, shown in the last row, so we had to limit ourselves to the differential equations for 25 master integrals  $\mathbf{J}=(J_1,\ldots,J_{25})^{T}$. They depend nontrivially on the dimensionless variable $x=t/s$.
The differential system has the form \eqref{eq:DEsystem} where the explicit form of the matrix $\matr M(\epsilon,x)$ is not presented here to save space and to avoid cluttering. There are three singular points of the system, $x=0,-1,\infty$. Note that these points correspond to the conditions $t=0$, $u=0$, and $s=0$, respectively. Nontrivial diagonal blocks of $\matr M$ have indices $\{9,10\}$, $\{11,12\}$, $\{16,17\}$, $\{18,19\}$, $\{20,21,22\}$, $\{23,24,25\}$. Let us explain how our algorithm works on the example of the block spanned by indices $\{23,24,25\}$. It has the form
\begin{equation}\label{eq:M_example}
\matr M_{\{23-25\}}(\epsilon,x)=
\matr A(\epsilon)/x+\matr B(\epsilon)/(x+1)\,,
\end{equation}
where
\begin{gather}
\matr A(\epsilon)=\left(
\begin{array}{ccc}
 -\epsilon -1 & 0 & -\frac{\epsilon +1}{5 \epsilon +1} \\
 2 (4 \epsilon +1) (5 \epsilon +1) & -3 \epsilon -1 & 2 (\epsilon +1) \\
 \frac{(2 \epsilon +1) (4 \epsilon +1) (5 \epsilon +1)}{\epsilon +1} & 0 & 5 \epsilon +1 \\
\end{array}
\right)\,,\quad
\matr B(\epsilon)=\left(
\begin{array}{ccc}
 3 \epsilon  & -\frac{\epsilon }{5 \epsilon +1} & \frac{\epsilon +1}{5 \epsilon +1} \\
 0 & -\epsilon -1 & -2 (\epsilon +1) \\
 0 & -\frac{\epsilon  (2 \epsilon +1)}{\epsilon +1} & -\epsilon  \\
\end{array}
\right)
\end{gather}
Since $\matr M_{\{23-25\}}(\epsilon,x)$ already has a Fuchsian form, we skip steps described in Algorithm \ref{alg:Fuchsian} and pass to the Algorithm \ref{alg:Normal}. From now on let us denote the matrix residue at infinity as $\matr C(\epsilon)$, 
\begin{equation}
\matr C(\epsilon)=-\matr A(\epsilon)-\matr B(\epsilon).
\end{equation}
The eigenvalues of the matrices $\matr A$, $\matr B$, and $\matr C$ are, respectively
\begin{equation}
\matr A\colon \{-3 \epsilon -1,\epsilon ,3 \epsilon \}  \,, \quad 
\matr B\colon\{3 \epsilon ,\epsilon ,-3 \epsilon -1\} \,, \quad 
\matr C\colon \{-4 \epsilon -1,1,2 \epsilon +2\}\,.
\end{equation} 
As it should be, the sum of all eigenvalues is zero. The right and  left eigenvectors of the matrices $\matr A$ and $\matr C$, corresponding to the eigenvalues $-3 \epsilon -1$ and $2 \epsilon +2$, respectively, are
\begin{equation}
u=(0, 1, 0)^T\,,\quad 
v^\dagger=(-2 (1 + 5 \epsilon), 1, 0)\,.
\end{equation}
Since $v^\dagger u=1\neq 0$, the points $x=0$ and $x=\infty$ can be mutually balanced via $\matr{\bal}(u v^\dagger,0,\infty,x)$. After the transformation we have the same form \eqref{eq:M_example} with
\begin{gather}
\matr A(\epsilon)=\left(
\begin{array}{ccc}
 \epsilon -1 & -\frac{\epsilon }{5 \epsilon +1} & \frac{-\epsilon -1}{5 \epsilon +1} \\
 40 \epsilon ^2-2 \epsilon -2 & -5 \epsilon  & -2 \epsilon -2 \\
 \frac{(2 \epsilon +1) (5 \epsilon +1) (6 \epsilon +1)}{\epsilon +1} & -\frac{\epsilon  (2 \epsilon +1)}{\epsilon +1} & 5 \epsilon +1 \\
\end{array}
\right)\,,\quad
\matr B(\epsilon)=\left(
\begin{array}{ccc}
 -\epsilon  & \frac{\epsilon }{5 \epsilon +1} & \frac{\epsilon +1}{5 \epsilon +1} \\
 20 \epsilon +4 & 3 \epsilon -1 & 6 \epsilon +6 \\
 -\frac{4 \epsilon  (2 \epsilon +1) (5 \epsilon +1)}{\epsilon +1} & \frac{\epsilon  (2 \epsilon +1)}{\epsilon +1} & -\epsilon  \\
\end{array}
\right)
\end{gather}
The eigenvalues of $\matr A$, $\matr B$, and $\matr C$ are now
\begin{equation}
\matr A\colon \{-3 \epsilon,\epsilon ,3 \epsilon \}  \,, \quad 
\matr B\colon\{3 \epsilon ,\epsilon ,-3 \epsilon -1\} \,, \quad 
\matr C\colon \{-4 \epsilon -1,1,2 \epsilon +1\}\,.
\end{equation} 
Note that a pair of eigenvalues has been shifted towards the interval $[-1/2,1/2)$. Now the right and  left eigenvectors of the matrices $\matr B$ and $\matr C$, corresponding to the eigenvalues $-3 \epsilon -1$ and $2 \epsilon +1$, respectively, are
\begin{equation}
u=\left(0,\epsilon +1,-\epsilon \right)^T\,,\quad 
v^\dagger=\left((5 \epsilon +1) (8 \epsilon +3),-3 \epsilon -1,2\epsilon +2\right)\,.
\end{equation}
Again, $v^\dagger u\neq 0$, therefore, we can mutually balance $x=-1$ and $x=\infty$  via $\matr{\bal}(u v^\dagger/(v^\dagger u),-1,\infty,x)$. After the transformation we have the form \eqref{eq:M_example} with

\begin{gather}
\matr A(\epsilon)=\left(
\begin{array}{ccc}
 \epsilon -1 & -\frac{\epsilon }{5 \epsilon +1} & -\frac{\epsilon +1}{5 \epsilon +1} \\
 2 (4 \epsilon -1) (5 \epsilon +1) & -5 \epsilon  & -2 (\epsilon +1) \\
 \frac{(2 \epsilon +1) (5 \epsilon +1) (6 \epsilon +1)}{\epsilon +1} & -\frac{\epsilon  (2 \epsilon +1)}{\epsilon +1} & 5 \epsilon +1 \\
\end{array}
\right)\,,\quad
\matr B(\epsilon)=\left(
\begin{array}{ccc}
 -\epsilon  & \frac{\epsilon }{5 \epsilon +1} & \frac{\epsilon +1}{5 \epsilon +1} \\
 -32 \epsilon ^2-6 \epsilon  & \frac{\epsilon  (21 \epsilon +4)}{5 \epsilon +1} & \frac{(\epsilon +1) (16 \epsilon +3)}{5 \epsilon +1} \\
 \frac{-88 \epsilon ^3-48 \epsilon ^2-6 \epsilon }{\epsilon +1} & \frac{\epsilon  \left(34 \epsilon ^2+17 \epsilon +2\right)}{(\epsilon +1) (5 \epsilon +1)} & \frac{-11 \epsilon ^2-2 \epsilon }{5 \epsilon +1} \\
\end{array}
\right)
\end{gather}
The eigenvalues of $\matr A$, $\matr B$, and $\matr C$ are
\begin{equation}
\matr A\colon \{-3 \epsilon,\epsilon ,3 \epsilon \}  \,, \quad 
\matr B\colon\{3 \epsilon ,\epsilon ,-3 \epsilon \} \,, \quad 
\matr C\colon \{-4 \epsilon -1,1,2 \epsilon\}\,.
\end{equation}
Now the system is normalized at $x=0$ and $x=-1$, but not in $x=\infty$. In order to normalize the system at all points, we need to perform intermediate transformation moving one unnormalized eigenvalue to another point. In particular, we may use the right and  left eigenvectors of the matrices $\matr C$ and $\matr B$, corresponding to the eigenvalues $-4 \epsilon -1$ and $\epsilon$, respectively, which are
\begin{equation}
u=\left(0,\epsilon +1,4 \epsilon +1\right)^T\,,\quad 
v^\dagger=\left(-16 \epsilon -3,1,0\right)\,,
\end{equation}
and make the transformation $\matr{\bal}(u v^\dagger/(v^\dagger u),\infty,-1,x)$. After the transformation we have 
\begin{gather}
\matr A(\epsilon)=\left(
\begin{array}{ccc}
 \epsilon -1 & -\frac{\epsilon }{5 \epsilon +1} & -\frac{\epsilon +1}{5 \epsilon +1} \\
 2 (4 \epsilon -1) (5 \epsilon +1) & -5 \epsilon  & -2 (\epsilon +1) \\
 \frac{(2 \epsilon +1) (5 \epsilon +1) (6 \epsilon +1)}{\epsilon +1} & -\frac{\epsilon  (2 \epsilon +1)}{\epsilon +1} & 5 \epsilon +1 \\
\end{array}
\right)\,,\quad
\matr B(\epsilon)=\left(
\begin{array}{ccc}
 3 (5 \epsilon +1) & -\frac{4 \epsilon +1}{5 \epsilon +1} & \frac{\epsilon +1}{5 \epsilon +1} \\
 2 (4 \epsilon +1) (19 \epsilon +4) & -7 \epsilon -3 & 2 (\epsilon +1) \\
 -\frac{(4 \epsilon +1) \left(118 \epsilon ^2+29 \epsilon +1\right)}{\epsilon +1} & \frac{8 \epsilon  (4 \epsilon +1)}{\epsilon +1} & -7 \epsilon -1 \\
\end{array}
\right)
\end{gather}
The eigenvalues of $\matr A$, $\matr B$, and $\matr C$ are
\begin{equation}
\matr A\colon \{-3 \epsilon,\epsilon ,3 \epsilon \}  \,, \quad 
\matr B\colon\{3 \epsilon ,\epsilon-1 ,-3 \epsilon \} \,, \quad 
\matr C\colon \{-4 \epsilon,1,2 \epsilon\}\,.
\end{equation}
Now it is easy to check that $x=-1$ and $x=\infty$ can be mutually balanced via $\matr{\bal}(u v^\dagger/(v^\dagger u),-1,\infty,x)$, where
\begin{equation}
u=\left(0,\epsilon +1,4 \epsilon +1\right)^T\,,\quad 
v^\dagger=\left(-2 (6 \epsilon +1),1,0\right)
\end{equation}
are the corresponding eigenvectors of $\matr B$ and $\matr C$.
After that we have
\begin{gather}
\matr A(\epsilon)=\left(
\begin{array}{ccc}
 \epsilon -1 & -\frac{\epsilon }{5 \epsilon +1} & -\frac{\epsilon +1}{5 \epsilon +1} \\
 2 (4 \epsilon -1) (5 \epsilon +1) & -5 \epsilon  & -2 (\epsilon +1) \\
 \frac{(2 \epsilon +1) (5 \epsilon +1) (6 \epsilon +1)}{\epsilon +1} & -\frac{\epsilon  (2 \epsilon +1)}{\epsilon +1} & 5 \epsilon +1 \\
\end{array}
\right)\,,\quad
\matr B(\epsilon)=\left(
\begin{array}{ccc}
 3 \epsilon +1 & \frac{\epsilon }{5 \epsilon +1} & \frac{\epsilon +1}{5 \epsilon +1} \\
 2 (2 \epsilon +1) (6 \epsilon +1) & \frac{\epsilon  (17 \epsilon +3)}{5 \epsilon +1} & \frac{2 (\epsilon +1) (6 \epsilon +1)}{5 \epsilon +1} \\
 -\frac{(3 \epsilon +1) (6 \epsilon +1) (8 \epsilon +1)}{\epsilon +1} & \frac{\epsilon  (3 \epsilon +1) (6 \epsilon +1)}{(\epsilon +1) (5 \epsilon +1)} & -\frac{27 \epsilon ^2+10 \epsilon +1}{5 \epsilon +1} \\
\end{array}
\right)
\end{gather}
with the eigenvalues
\begin{equation}
\matr A\colon \{-3 \epsilon,\epsilon ,3 \epsilon \}  \,, \quad 
\matr B\colon\{3 \epsilon ,\epsilon ,-3 \epsilon \} \,, \quad 
\matr C\colon \{-4 \epsilon,0,2 \epsilon\}\,.
\end{equation}
At this stage we have succeeded to normalize all matrix residues $\matr A$, $\matr B$, and  $\matr C$. Finally, we solve the system of linear equations
\begin{equation}
\frac{\matr{A}(\epsilon)}{\epsilon}\matr T=\matr T\frac{\matr{A}(\mu)}{\mu}\,,\quad
\frac{\matr{B}(\epsilon)}{\epsilon}\matr T=\matr T\frac{\matr{B}(\mu)}{\mu}
\end{equation}
with respect to the matrix elements of $\matr T$. We obtain
\begin{equation}
\matr T(\epsilon,\mu)=\left(
\begin{array}{ccc}
 (\epsilon +1) \mu  (5 \mu +1) & 0 & 0 \\
 -2 (\epsilon +1) (\epsilon -\mu ) (5 \mu +1) & \epsilon  (\epsilon +1) (5 \mu +1) & 0 \\
 (7 \epsilon +1) (\epsilon -\mu ) (5 \mu +1) & -\epsilon  (\epsilon -\mu ) & \epsilon  (5 \epsilon +1) (\mu +1) \\
\end{array}
\right)
\end{equation}
up to an arbitrary factor. We can now put $\mu$ to any constant number provided $\matr T$ remains invertible (in particular, we can not put $\mu$ to $0$, $-1$, or $-1/5$). We choose $\mu=1$. Making the transformation with $\matr T(\epsilon,1)$ we finally obtain the desired $\epsilon$-form:
\begin{equation}
\matr M_{\{23-25\}}(\epsilon,x)=\epsilon \left(
\begin{array}{ccc}
 \frac{4}{x+1} & -\frac{1}{6 x (x+1)} & -\frac{1}{3 x (x+1)} \\
 \frac{6 (13 x+6)}{x (x+1)} & -\frac{5 (x+3)}{3 x (x+1)} & \frac{2 (x-6)}{3 x (x+1)} \\
 -\frac{63 (x-1)}{x (x+1)} & \frac{5 x-9}{6 x (x+1)} & -\frac{x-18}{3 x (x+1)} \\
\end{array}
\right)
\end{equation}
At this stage one may want to make yet another transformation with a constant matrix, which reduces one of the matrix residues to diagonal form. E.g., we can take the matrix, transforming $\matr A$ to diagonal form
\begin{equation}
\matr T=\left(
\begin{array}{ccc}
 1 & 1 & 1 \\
 24 & 12 & 12 \\
 -3 & -15 & -9 \\
\end{array}
\right)\,.
\end{equation}
The resulting matrix has a somewhat simpler form:
\begin{equation}
\matr M_{\{23-25\}}(\epsilon,x)=\epsilon \left(
\begin{array}{ccc}
 -\frac{x+3}{x (x+1)} & 0 & \frac{1}{3 (x+1)} \\
 0 & \frac{2 x+3}{x (x+1)} & \frac{8}{3 (x+1)} \\
 \frac{5}{x+1} & \frac{2}{x+1} & \frac{1}{x} \\
\end{array}
\right)\,.
\end{equation}

In a similar way we reduce all diagonal blocks to $\epsilon$-form. Finally, using the approach of Section \ref{sec:btform}, we obtain the system
\begin{equation}
\partial_{x}\widetilde{\mathbf{J}}=\epsilon\left[\frac{\matr{S}_1}{x}+\frac{\matr{S}_2}{x+1}\right]\widetilde{\mathbf{J}}\,,
\end{equation}
where $\matr S_1$ and $\matr S_2$ are presented in the appendix. To avoid clutter, we do not present here the transformation matrix $\matr T$. Both this matrix and the original form of the system are available upon request from the author.

\section{Conclusion}

We have presented a practical algorithm of the reduction of differential system to $\epsilon$-form. The main tool of our approach is the transformation \eqref{eq:balance} which we call balance. We have shown how to construct a balance which does not increase the Poincar\'e rank of the system at any point on the extended complex plane. Moreover, we have shown how to construct the balances which can be used to lower the Poincar\'e rank $p$ at the point with $p>0$ and to normalize the eigenvalues of the matrix residue at the point with $p=0$. The reduction to $\epsilon$-form can be divided into three stages
\begin{itemize}
\item[1.] Reduction to Fuchsian form, Algorithm \ref{alg:Fuchsian}.
\item[2.] Normalizing eigenvalues, Algorithm \ref{alg:Normal}.
\item[3.] Factoring out $\epsilon$, Section \ref{sec:factoring_epsilon}. 
\end{itemize}
We have also shown how to use the block-triangular form of the system to alleviate computation. Namely, we first apply the above three step to each diagonal block and find the corresponding matrices $\matr T_i$ transforming each block to $\epsilon$-form. After the block-diagonal transformation $\matr T=\diag(\matr T_1,\matr T_2,\ldots)$ the diagonal blocks of the transformed system are in $\epsilon$-form. Then we use prescriptions of Section \ref{sec:btform} and, finally, factor out $\epsilon$ from the whole system. The latter can be done in such a way as to preserve the block-diagonal structure of the system, as explained in the end of Section \ref{sec:btform}.

There may be obstructions to the construction of the appropriate balance due to the orthogonality of the left and right eigenvectors. However, the appearance of obstructions is expected due to the negative solution of the 21st Hilbert problem by Bolibrukh \cite{Bolibrukh1989}. For a Fuchsian system with normalized eigenvalues we have shown how to find the constant transformation reducing the system to $\epsilon$-form. We have successfully applied our method to the reduction of several differential systems. We have also checked that for the case of three-loop all-massive sunrise propagator master integrals the obstruction to the reduction appears. This obstruction naturally corresponds to the fact that these master integrals can not be expressed in terms of harmonic polylogarithms \cite{BlochKerrVanhove2014}. 

The example presented in Section \ref{sec:example} did not require the reduction of the system to Fuchsian form, as described by Algorithm \ref{alg:Fuchsian},  since all diagonal blocks have been already in Fuchsian form. Though it may be considered as a poor choice of the example, we underline, that the reduction to a Fuchsian form can, in principle, be done solely by means of the Barkatou\&Pfl\"ugel algorithm \cite{BarkatouPfluegel2007,BarkatouPfluegel2009}. Thus, a demonstration of the viability of our algorithm for this stage is not very crucial. On the other hand, the system \eqref{eq:M_example} is not of the form assumed in Refs. \cite{ArgeriDiMastroliaMirabellaSchlenkothers2014,GehrmannManteuffelTancrediWeihs2014} and, therefore, its reduction to $\epsilon$-form with the tools developed in the present paper seems to be quite expository. 

Finally, we note that, though it is possible to make the reduction manually, it is very desirable to automatize the process as much as possible. A dedicated \texttt{Mathematica} package is being developed now and will be presented elsewhere.

\paragraph{Acknowledgments.}
The work  has been supported in part by the Ministry of Education and Science of the Russian Federation and the RFBR grants nos. 13-02-01023 and 15-02-07893. I am grateful to Thomas Gehrmann, Johannes Henn, and Andrei Pomeransky for the interest to the work and useful discussions. Special thanks go to Andrei Pomeransky for pointing out Ref. \cite{zakharov1980soliton}, which triggered the idea of using balances for the reduction. I'm grateful to Vladimir Smirnov for pointing out some typos in the preliminary version of the paper. I appreciate kind hospitality of the  Physics Department of Z\"urich University where this work has been finished.
\paragraph{Note added in proof.} After this paper has been finished, lecture notes on differential equations method by Henn  \cite{Henn2014} have been published. These lecture notes contain extended review of the approach of Ref. \cite{Henn2013}. In particular, the choice of the integrals with homogeneous transcendental weight is discussed in detail.

\appendix
\section*{Appendix. The form of matrices $\matr S_1$ and $\matr S_2$.}
\begin{align*}
\matr S_1&=
\left(
\mbox{\tiny$\begin{array}{ccccccccccccccccccccccccc}
 0 & 0 & 0 & 0 & 0 & 0 & 0 & 0 & 0 & 0 & 0 & 0 & 0 & 0 & 0 & 0 & 0 & 0 & 0 & 0 & 0 & 0 & 0 & 0 & 0 \\
 0 & -3 & 0 & 0 & 0 & 0 & 0 & 0 & 0 & 0 & 0 & 0 & 0 & 0 & 0 & 0 & 0 & 0 & 0 & 0 & 0 & 0 & 0 & 0 & 0 \\
 0 & 0 & 0 & 0 & 0 & 0 & 0 & 0 & 0 & 0 & 0 & 0 & 0 & 0 & 0 & 0 & 0 & 0 & 0 & 0 & 0 & 0 & 0 & 0 & 0 \\
 0 & 0 & 0 & 0 & 0 & 0 & 0 & 0 & 0 & 0 & 0 & 0 & 0 & 0 & 0 & 0 & 0 & 0 & 0 & 0 & 0 & 0 & 0 & 0 & 0 \\
 0 & 0 & 0 & 0 & 0 & 0 & 0 & 0 & 0 & 0 & 0 & 0 & 0 & 0 & 0 & 0 & 0 & 0 & 0 & 0 & 0 & 0 & 0 & 0 & 0 \\
 0 & 2 & 0 & 0 & 0 & -3 & 0 & 0 & 0 & 0 & 0 & 0 & 0 & 0 & 0 & 0 & 0 & 0 & 0 & 0 & 0 & 0 & 0 & 0 & 0 \\
 0 & 0 & 0 & 0 & 0 & 0 & 2 & 0 & 0 & 0 & 0 & 0 & 0 & 0 & 0 & 0 & 0 & 0 & 0 & 0 & 0 & 0 & 0 & 0 & 0 \\
 0 & -2 & 0 & 0 & 0 & 0 & 0 & -3 & 0 & 0 & 0 & 0 & 0 & 0 & 0 & 0 & 0 & 0 & 0 & 0 & 0 & 0 & 0 & 0 & 0 \\
 0 & 0 & 0 & 0 & 0 & 0 & 0 & 0 & 1 & 0 & 0 & 0 & 0 & 0 & 0 & 0 & 0 & 0 & 0 & 0 & 0 & 0 & 0 & 0 & 0 \\
 0 & 0 & 0 & 0 & 0 & 0 & 0 & 0 & 0 & 0 & 0 & 0 & 0 & 0 & 0 & 0 & 0 & 0 & 0 & 0 & 0 & 0 & 0 & 0 & 0 \\
 0 & -6 & 0 & 0 & 0 & 0 & 0 & 0 & 0 & 0 & -3 & 0 & 0 & 0 & 0 & 0 & 0 & 0 & 0 & 0 & 0 & 0 & 0 & 0 & 0 \\
 0 & 0 & 0 & 0 & 0 & 0 & 0 & 0 & 0 & 0 & 0 & 0 & 0 & 0 & 0 & 0 & 0 & 0 & 0 & 0 & 0 & 0 & 0 & 0 & 0 \\
 0 & 0 & 0 & 0 & 0 & 0 & 0 & 0 & 0 & 0 & 0 & 0 & 3 & 0 & 0 & 0 & 0 & 0 & 0 & 0 & 0 & 0 & 0 & 0 & 0 \\
 0 & -2 & 0 & 0 & 0 & 0 & 0 & 0 & 0 & 0 & 0 & 0 & 0 & -3 & 0 & 0 & 0 & 0 & 0 & 0 & 0 & 0 & 0 & 0 & 0 \\
 0 & 0 & 0 & 0 & 0 & 0 & 0 & 0 & 0 & 0 & 0 & 0 & 0 & 0 & 0 & 0 & 0 & 0 & 0 & 0 & 0 & 0 & 0 & 0 & 0 \\
 0 & 0 & 0 & 0 & 0 & 0 & 0 & 0 & 0 & 0 & 0 & 0 & 0 & 0 & 0 & 1 & 0 & 0 & 0 & 0 & 0 & 0 & 0 & 0 & 0 \\
 0 & 0 & 0 & 0 & 0 & 0 & 0 & 0 & 0 & 0 & 0 & 0 & 0 & 0 & 0 & 0 & 0 & 0 & 0 & 0 & 0 & 0 & 0 & 0 & 0 \\
 0 & 0 & 0 & 0 & 0 & 0 & 0 & 0 & 0 & 0 & 0 & 0 & 0 & 0 & 0 & 0 & 0 & -3 & 0 & 0 & 0 & 0 & 0 & 0 & 0 \\
 0 & 0 & 0 & 0 & 0 & 0 & 0 & 0 & 0 & 0 & 0 & 0 & 0 & 0 & 0 & 0 & 0 & 0 & 3 & 0 & 0 & 0 & 0 & 0 & 0 \\
 0 & 0 & 0 & 0 & 0 & -6 & 0 & -12 & 0 & 0 & 2 & 0 & 0 & 0 & 0 & 0 & 0 & 0 & 0 & -3 & 0 & 0 & 0 & 0 & 0 \\
 0 & 0 & 0 & 0 & 0 & 0 & 0 & 0 & 0 & 0 & 0 & 0 & 0 & 0 & 0 & 0 & 0 & 0 & 0 & 0 & 1 & 0 & 0 & 0 & 0 \\
 0 & 0 & 0 & 0 & 0 & 0 & 0 & 0 & 0 & 0 & 0 & 0 & 0 & 0 & 0 & 0 & 0 & 0 & 0 & 0 & 0 & 0 & 0 & 0 & 0 \\
 0 & 0 & 0 & 0 & 0 & \frac{1}{2} & 0 & 0 & 0 & 0 & 0 & 0 & 0 & \frac{1}{2} & 0 & 0 & 0 & 0 & 0 & 0 & 0 & 0 & -3 & 0 & 0 \\
 0 & 0 & 0 & 0 & 0 & 0 & 0 & 0 & 0 & 0 & 0 & 0 & 0 & 0 & 0 & 0 & 0 & 0 & 0 & 0 & 0 & 0 & 0 & 3 & 0 \\
 0 & 0 & 0 & 0 & 0 & 0 & 0 & 0 & 0 & 0 & 0 & 0 & 0 & 0 & 0 & 0 & 0 & 0 & 0 & 0 & 0 & 0 & 0 & 0 & 1 \\
\end{array}$}
\right)\,,\\
\matr S_2&=\left(\mbox{\tiny
$\begin{array}{ccccccccccccccccccccccccc}
 -3 & 0 & 0 & 0 & 0 & 0 & 0 & 0 & 0 & 0 & 0 & 0 & 0 & 0 & 0 & 0 & 0 & 0 & 0 & 0 & 0 & 0 & 0 & 0 & 0 \\
 0 & 0 & 0 & 0 & 0 & 0 & 0 & 0 & 0 & 0 & 0 & 0 & 0 & 0 & 0 & 0 & 0 & 0 & 0 & 0 & 0 & 0 & 0 & 0 & 0 \\
 0 & 0 & 0 & 0 & 0 & 0 & 0 & 0 & 0 & 0 & 0 & 0 & 0 & 0 & 0 & 0 & 0 & 0 & 0 & 0 & 0 & 0 & 0 & 0 & 0 \\
 0 & 0 & 0 & 0 & 0 & 0 & 0 & 0 & 0 & 0 & 0 & 0 & 0 & 0 & 0 & 0 & 0 & 0 & 0 & 0 & 0 & 0 & 0 & 0 & 0 \\
 0 & 0 & 0 & 0 & 0 & 0 & 0 & 0 & 0 & 0 & 0 & 0 & 0 & 0 & 0 & 0 & 0 & 0 & 0 & 0 & 0 & 0 & 0 & 0 & 0 \\
 2 & 0 & 0 & 0 & 0 & -3 & 0 & 0 & 0 & 0 & 0 & 0 & 0 & 0 & 0 & 0 & 0 & 0 & 0 & 0 & 0 & 0 & 0 & 0 & 0 \\
 -2 & 0 & 0 & 1 & 0 & 0 & -3 & 0 & 0 & 0 & 0 & 0 & 0 & 0 & 0 & 0 & 0 & 0 & 0 & 0 & 0 & 0 & 0 & 0 & 0 \\
 0 & 0 & 0 & \frac{2}{3} & 0 & 0 & 0 & 2 & 0 & 0 & 0 & 0 & 0 & 0 & 0 & 0 & 0 & 0 & 0 & 0 & 0 & 0 & 0 & 0 & 0 \\
 0 & 0 & 0 & 0 & 0 & 0 & 0 & 0 & -4 & -2 & 0 & 0 & 0 & 0 & 0 & 0 & 0 & 0 & 0 & 0 & 0 & 0 & 0 & 0 & 0 \\
 -6 & 0 & 0 & 0 & 0 & 0 & 0 & 0 & 2 & 1 & 0 & 0 & 0 & 0 & 0 & 0 & 0 & 0 & 0 & 0 & 0 & 0 & 0 & 0 & 0 \\
 0 & 0 & 0 & 0 & 0 & 0 & 0 & 0 & 0 & 0 & \frac{4}{3} & -\frac{2}{3} & 0 & 0 & 0 & 0 & 0 & 0 & 0 & 0 & 0 & 0 & 0 & 0 & 0 \\
 0 & 0 & 0 & 0 & 0 & 0 & 0 & 0 & 0 & 0 & \frac{2}{3} & -\frac{1}{3} & 0 & 0 & 0 & 0 & 0 & 0 & 0 & 0 & 0 & 0 & 0 & 0 & 0 \\
 2 & 0 & -2 & 0 & 0 & 0 & 0 & 0 & 0 & 0 & 0 & 0 & -3 & 0 & 0 & 0 & 0 & 0 & 0 & 0 & 0 & 0 & 0 & 0 & 0 \\
 0 & 0 & 2 & 0 & 0 & 0 & 0 & 0 & 0 & 0 & 0 & 0 & 0 & 3 & 0 & 0 & 0 & 0 & 0 & 0 & 0 & 0 & 0 & 0 & 0 \\
 0 & 0 & 0 & 0 & 0 & 0 & 0 & 0 & 0 & 0 & 0 & 0 & 0 & 0 & 0 & 0 & 0 & 0 & 0 & 0 & 0 & 0 & 0 & 0 & 0 \\
 0 & 0 & 0 & 0 & 0 & 0 & 36 & 0 & 0 & 0 & 0 & 0 & 0 & 0 & 0 & -9 & -6 & 0 & 0 & 0 & 0 & 0 & 0 & 0 & 0 \\
 -36 & 0 & 0 & 18 & -3 & 0 & -72 & 0 & 0 & 0 & 0 & 0 & 0 & 0 & 0 & 12 & 9 & 0 & 0 & 0 & 0 & 0 & 0 & 0 & 0 \\
 0 & 0 & 0 & -4 & 4 & 0 & 0 & 12 & 0 & 0 & 0 & 0 & 0 & 0 & 0 & 0 & 0 & 2 & -2 & 0 & 0 & 0 & 0 & 0 & 0 \\
 0 & 0 & 0 & -4 & 2 & 0 & 0 & 0 & 0 & 0 & 0 & 0 & 0 & 0 & 0 & 0 & 0 & 1 & -1 & 0 & 0 & 0 & 0 & 0 & 0 \\
 0 & 0 & 0 & 0 & 0 & 0 & -12 & 0 & 0 & 0 & \frac{3}{25} & 0 & 0 & 0 & 0 & 0 & 0 & 0 & 0 & \frac{7}{6} & -\frac{1}{2} & \frac{1}{3} & 0 & 0 & 0 \\
 -\frac{24}{25} & \frac{80}{7} & 0 & -4 & 0 & -\frac{78}{5} & -36 & -36 & 16 & 16 & \frac{127}{15} & \frac{158}{75} & 0 & 0 & 0 & 0 & 0 & 0 & 0 & -\frac{5}{2} & -\frac{7}{2} & 1 & 0 & 0 & 0 \\
 -\frac{306}{5} & \frac{60}{7} & 0 & 2 & 0 & -\frac{54}{5} & -48 & -48 & 12 & 12 & \frac{658}{75} & \frac{22}{75} & 0 & 0 & 0 & 0 & 0 & 0 & 0 & -\frac{10}{3} & -2 & \frac{1}{3} & 0 & 0 & 0 \\
 0 & 0 & 0 & 0 & -\frac{11}{27} & -\frac{5}{6} & \frac{1}{2} & -\frac{29}{36} & 0 & 0 & 0 & 0 & -\frac{1}{2} & \frac{17}{72} & \frac{5}{36} & 0 & -\frac{1}{36} & \frac{1}{24} & \frac{29}{216} & 0 & 0 & 0 & 2 & 0 & \frac{1}{3} \\
 0 & -\frac{10}{9} & -\frac{65}{9} & 0 & \frac{5}{9} & \frac{5}{3} & -10 & 0 & 0 & 0 & 0 & 0 & -\frac{25}{3} & \frac{5}{3} & \frac{5}{9} & \frac{4}{3} & \frac{5}{9} & \frac{5}{9} & 0 & 0 & 0 & 0 & 0 & -1 & \frac{8}{3} \\
 0 & \frac{35}{36} & \frac{5}{36} & \frac{5}{18} & \frac{5}{36} & -\frac{13}{6} & -\frac{27}{2} & \frac{49}{72} & 0 & 0 & 0 & 0 & -\frac{35}{6} & -\frac{145}{72} & -\frac{5}{36} & 2 & \frac{43}{36} & \frac{145}{432} & -\frac{25}{36} & 0 & 0 & 0 & 5 & 2 & 0 \\
\end{array}$}
\right)\,.
\end{align*}

\bibliographystyle{JHEP}
\providecommand{\href}[2]{#2}\begingroup\raggedright\endgroup

\end{document}